\documentclass[iop]{emulateapj}
\pdfoutput=1
\usepackage{graphicx}
\usepackage{color}

\begin{document}

\shortauthors{Rappaport et al.}
\shorttitle{Disintegrating Planet}
\title{Possible Disintegrating Short-Period Super-Mercury Orbiting KIC 12557548}

\author{S. Rappaport\altaffilmark{1}, A. Levine\altaffilmark{2}, E. Chiang\altaffilmark{3,4}, I. El Mellah\altaffilmark{1,5}, J. Jenkins\altaffilmark{6},  B. Kalomeni\altaffilmark{2,7}, E.~S.~Kite\altaffilmark{4,8}, 
M. Kotson\altaffilmark{1}, L. Nelson\altaffilmark{9}, L. Rousseau-Nepton\altaffilmark{10}, and K. Tran\altaffilmark{1}}

\altaffiltext{1}{37-602B, M.I.T. Department of Physics and Kavli
 Institute for Astrophysics and Space Research, 70 Vassar St.,
 Cambridge, MA, 02139; sar@mit.edu} 
 \altaffiltext{2}{37-575, M.I.T. Kavli
 Institute for Astrophysics and Space Research, 70 Vassar St.,
 Cambridge, MA, 02139; aml@space.mit.edu} 
 \altaffiltext{3}{Department of Astronomy, UC Berkeley, Hearst Field Annex B-20, Berkeley CA 94720-3411; echiang@astro.berkeley.edu} 
 \altaffiltext{4}{Department of Earth and Planetary Science, UC Berkeley, 307 McCone Hall, Berkeley CA 94720-4767} 
 \altaffiltext{5}{ENS Cachan, 61 avenue du Pr\'esident Wilson, 94235 Cachan, France; ielmelah@ens-cachan.fr}
  \altaffiltext{6}{SETI Institute/NASA Ames Research Center, Moffett Field, CA 
 94035; Jon.M.Jenkins@nasa.gov} 
 \altaffiltext{7}{Department of Astronomy and Space Sciences, University of Ege, 35100 Bornova-Izmir, Turkey; Department of Physics, Izmir Institute of Technology, Turkey}
\altaffiltext{8}{Division of Geological and Planetary Sciences, Caltech MC 150-21, Pasadena CA 91125; ekite@caltech.edu}
 \altaffiltext{9}{Department of Physics, Bishop's University, 2600 College St., Sherbrooke, QC J1M 1Z7; lnelson@ubishops.ca} 
\altaffiltext{10}{D\'epartement de physique, de g\'enie physique et d'optique
Universit\'e Laval, Qu\'ebec, QC G1K 7P4; laurie.r-nepton.1@ulaval.ca}

  \begin{abstract}
    We report here on the discovery of stellar occultations, observed
    with {\em Kepler}, that recur periodically at 15.685 hour
    intervals, but which vary in depth from a maximum of 1.3\% to a
    minimum that can be less than 0.2\%.  The star that is apparently
    being occulted is KIC 12557548, a V = 16 magnitude K dwarf with
    $T_{\rm eff,s} \simeq 4400$ K.  The out-of-occultation behavior
    shows no evidence for ellipsoidal light variations, indicating
    that the mass of the orbiting object is less than $\sim$$3\,M_{\rm
      J}$ (for an orbital period of 15.7 hr).  Because the eclipse 
      depths are highly variable, they
    cannot be due solely to transits of a single planet with a fixed
    size.  We discuss but dismiss a scenario involving a binary giant
    planet whose mutual orbit plane precesses, bringing one of the
    planets into and out of a grazing transit.  This scenario seems
    ruled out by the dynamical instability that would result from such
    a configuration.  We also briefly consider an eclipsing binary, possibly 
    containing an accretion disk, that either orbits KIC 12557548 in a 
    hierarchical triple configuration or is nearby on the sky, but we find 
    such a scenario inadequate to reproduce the observations.  
    The much more likely explanation---but one which still requires
    more quantitative development---involves macroscopic particles
    escaping the atmosphere of a slowly disintegrating planet not much
    larger than Mercury in size.  The particles could take the form of
    micron-sized pyroxene or aluminum oxide dust grains. 
The planetary surface is hot enough to sublimate and create a high-Z
    atmosphere; this atmosphere may be loaded with dust via cloud
    condensation or explosive volcanism. Atmospheric gas escapes
    the planet via a Parker-type thermal wind, dragging dust
    grains with it. 
We infer a mass loss rate from the observations of
    order 1~$M_\oplus/$Gyr, with a dust-to-gas ratio possibly of order
    unity. For our fiducial $0.1 M_{\oplus}$ planet (twice the mass of
    Mercury), the evaporation timescale may be $\sim$0.2 Gyr.  Smaller
    mass planets are disfavored because they evaporate still more
    quickly, as are larger mass planets because they have surface
    gravities too strong to sustain outflows with the requisite
    mass-loss rates.  The occultation profile evinces an
    ingress-egress asymmetry that could reflect a comet-like dust tail
    trailing the planet; we present simulations of such a tail.
\end{abstract}


\section{Introduction}
\label{sec:intro}

\begin{figure}
\begin{center}
\includegraphics[width=0.98 \columnwidth]{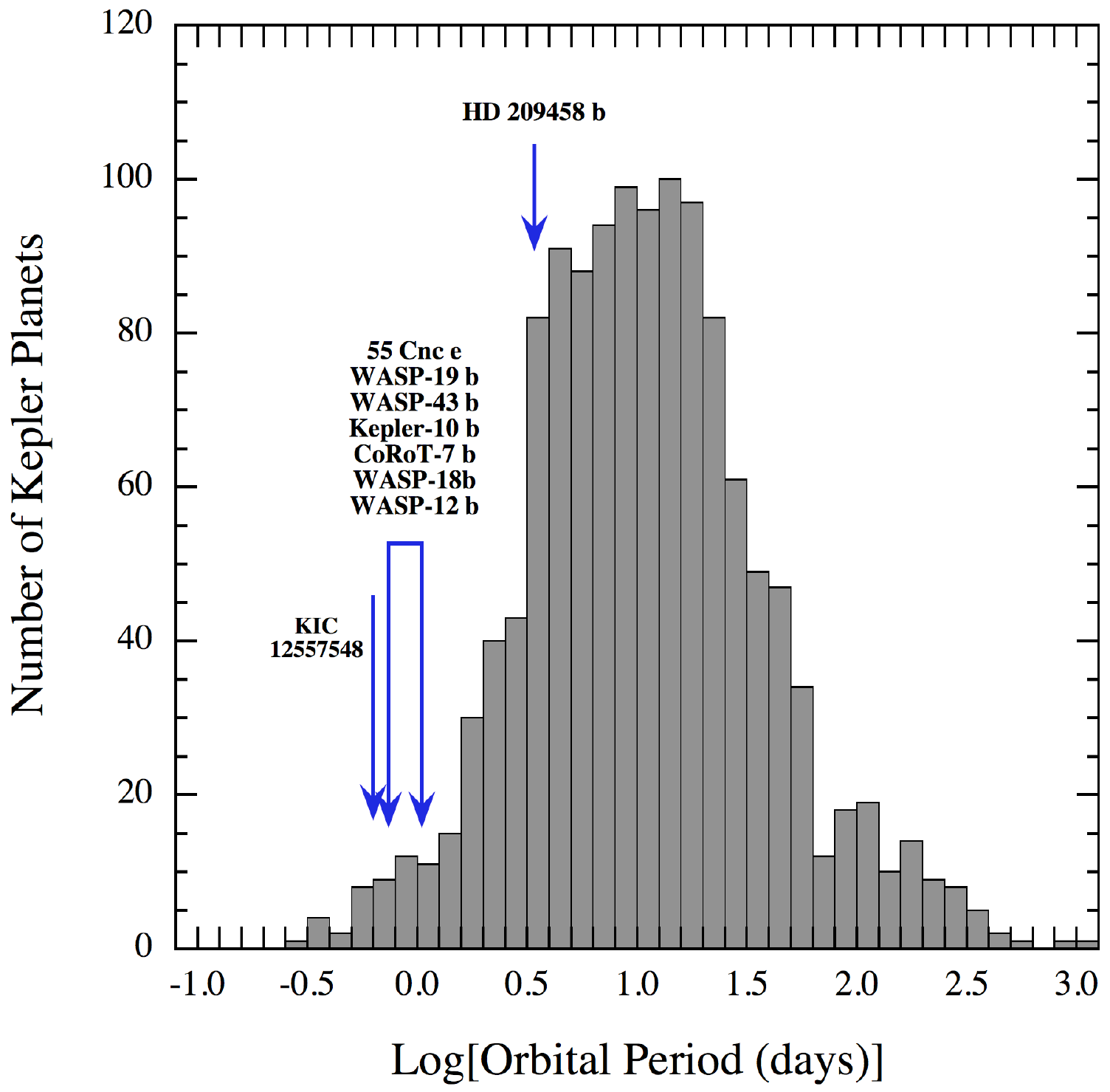}
\caption{Distribution of orbital periods among the exoplanet candidates discovered with {\em Kepler}.  The shorter period planets specifically named are not necessarily from the {\em Kepler} survey and are taken from the on-line catalog ``The Exoplanet Orbit Database'' (see text).}
\label{fig:porb_dist}
\end{center}
\end{figure}

The {\em Kepler} mission has now discovered some $\sim$1250 very good
exoplanet candidates via planetary transits of the parent star
(Borucki et al.~2011).  The transit depths range from a few percent
for giant gas planets to $\lesssim 10^{-4}$ for smaller planets
that are of the size of the Earth.  The distribution of orbital
periods for these systems is shown in Fig.~\ref{fig:porb_dist}.  The bulk of 
the systems have periods between $\sim$3 and 30
    days. As observations continue to be accumulated and analyzed, the
    newly discovered candidates will tend to be longer period systems;
    the short-period end of the period distribution will change
    relatively slowly. There are relatively few systems with periods
    below a few days, and furthermore, most of the systems shown in
    Fig.\,\ref{fig:porb_dist} with orbital periods less than a day are
    either now considered to be false positives (i.e., they are most
    likely associated with background blended binaries) or have not
    yet been followed up in the detail required for conclusive
    categorization (Borucki et al. 2011).  The confirmed short-period
    planets, most of which were not discovered by {\em
    Kepler},\footnote{Taken from the ``Exoplanet Orbit Database'' at
    http://exoplanets.org, produced by Jason Wright, Geoff Marcy, and
    the California Planet Survey consortium.} have orbital periods
    between 0.73 and 1.09 days. These orbital periods are indicated by
    dual arrows in Fig.~\ref{fig:porb_dist}.

The shortest planetary orbital period that has been reported is 17.8 hours 
   (Winn et al.~2011).\footnote{Only just recently an exoplanet with an even
   shorter orbital period, KOI 961.02, with $P_{\rm orb} = 10.9$ hours, was reported
   (Muirhead et al.~2012).}  The planet 55 Cnc e, the
   innermost of 5 planets revolving around 55 Cnc, was discovered with
   ground-based optical Doppler measurements (Fischer et al.~2008;
   Dawson \& Fabrycky 2010) and detected in transit by extensive
   satellite observations with the {\em MOST} satellite (Winn et
   al.~2011).  Winn et al.~(2011) estimate its mass and radius to be
   $\sim$$9\,M_\oplus$ and $2\,R_\oplus$. Planet orbits considerably
   shorter than a day are, in principle, quite possible, and their
   detection via transits using {\em Kepler} should not be problematic
   in a signal-to-noise sense.  There are at least two reasons,
   however, why such close-in, short-period planets have not been heretofore
   emphasized.  First, there are many short-period stellar binaries
   that, when their light is highly diluted by another star within a
   few arc seconds (e.g., the target star under
   observation by {\em Kepler}), are hard to distinguish from stars
   with transiting planets. Second, giant planets are expected to be
   destroyed by Roche lobe overflow if they are heated too strongly
   by their parent stars (Batygin, Stevenson, \& Bodenheimer 2011).

With {\em Kepler} data, the orbital period $P_{\rm orb}$ of a planet about a star can be measured. If the orbit is circular, and the planet has a mass $M_p$ much less than the mass of the parent star, then the size of its Roche lobe can be determined without knowing the mass of the parent star:
\begin{eqnarray}
R_L \simeq 1.95 \left(\frac{M_p}{M_J}\right)^{1/3} \left(\frac{P_{\rm orb}}{{\rm days}}\right)^{2/3}~R_J
\label{eqn:RL}
\end{eqnarray}
where $M_J$ and $R_J$ are the mass and radius of Jupiter,
respectively.  Therefore, even if the planet does not evaporate
because of the strong stellar radiation field, by the time a
Jupiter-mass planet 
migrates to an orbit with $P_{\rm orb}
\lesssim$ 9 hours, it will start to empty its envelope via Roche-lobe
overflow.

In this work we discuss the signature of a planet in a 15.685-hour
orbit about a K star.  The object came to our attention via periodic
occultations of the target star KIC 12557548.  
The term ``occultation'' is perhaps more appropriate than ``transit'' (although
we will use both terms interchangeably in this paper), because of the
fact that the depths of the events are highly time dependent.  As we
discuss in this work, the occultations are periodic to 1 part in
$10^5$ or more, and therefore must be due to the presence of an
orbiting companion.  However, the variation of the eclipse depths
rules out transits by a single opaque body.  More likely, we argue,
the occultations are due to a stream of dust particles resulting from
the slow disintegration of the planet.
	
In \S \ref{sec:LC}--\ref{sec:OMM} of this work we describe detailed
   analyses of the {\em Kepler} observations of KIC 12557548 from
   quarters Q2 through Q6.  In these sections we present light curves
   as well as power density spectra of the light curves.  We also
   present an optical spectrum which basically confirms the spectral
   properties of KIC 12557548 given in the Kepler Input Catalog (KIC).  In
   \S \ref{sec:interpret} we outline three explanations
   for the variable occultation depths, including grazing transits due
   to a pair of planets; an eclipsing binary, possibly 
    containing an accretion disk, that orbits KIC 12557548 in a hierarchical 
    triple configuration; or, more likely, obscuration by a stream of debris
   from a disintegrating planet.  We explore in \S \ref{sec:dust} the
   hypothesis of an occulting dust cloud in some depth.  In
   particular, we discuss how dust can be stripped off a 
   planet with substantial gravity.  We go on to compute the
   approximate shape of a dust stream that has left the planet and is
   shaped by radiation pressure.  Our findings are
   summarized in \S \ref{sec:summary}; there we also suggest future
   observations that may robustly establish the nature of the variable
   occultation depths.

\section{Data Analysis}

\subsection{Discovery}
\label{sec:discover}

The occultations of KIC 12557548 were discovered as part of a
   systematic search through the {\em Kepler} data base for eclipsing
   binaries that are not in the Prsa et al.~(2011) catalog.  We used
   publicly available quarter 2 (``Q2'') data from each of the
   $\sim$150,000 {\em Kepler} targets to carry out a general Fourier
   transform search for periodicities (Kotson et al.~2012).  The
   criteria for declaring a particular power density spectrum (PDS) as ``interesting'' are that
   (i) there is at least one peak with amplitude greater than 5
   standard deviations above the local mean of the spectrum, and (ii) there
   is at least one harmonic or subharmonic of the largest peak that is
   at least 3 standard deviations above the local mean.
   
In all, we find that $\sim$4800 power spectra contain interesting content, as
   defined above.  Of these, approximately 1800 are binaries that are
   already in the Prsa et al.~(2011) catalog, as well as some 300
   binaries, mostly of the contact variety, that appear to have not
   made it into that catalog (Kotson et al.~2012).  Most of the
   remaining FFTs of interest involve stars that exhibit pulsations of
   one sort or another.

\begin{figure*}
\begin{center}
\includegraphics[width=0.91 \textwidth]{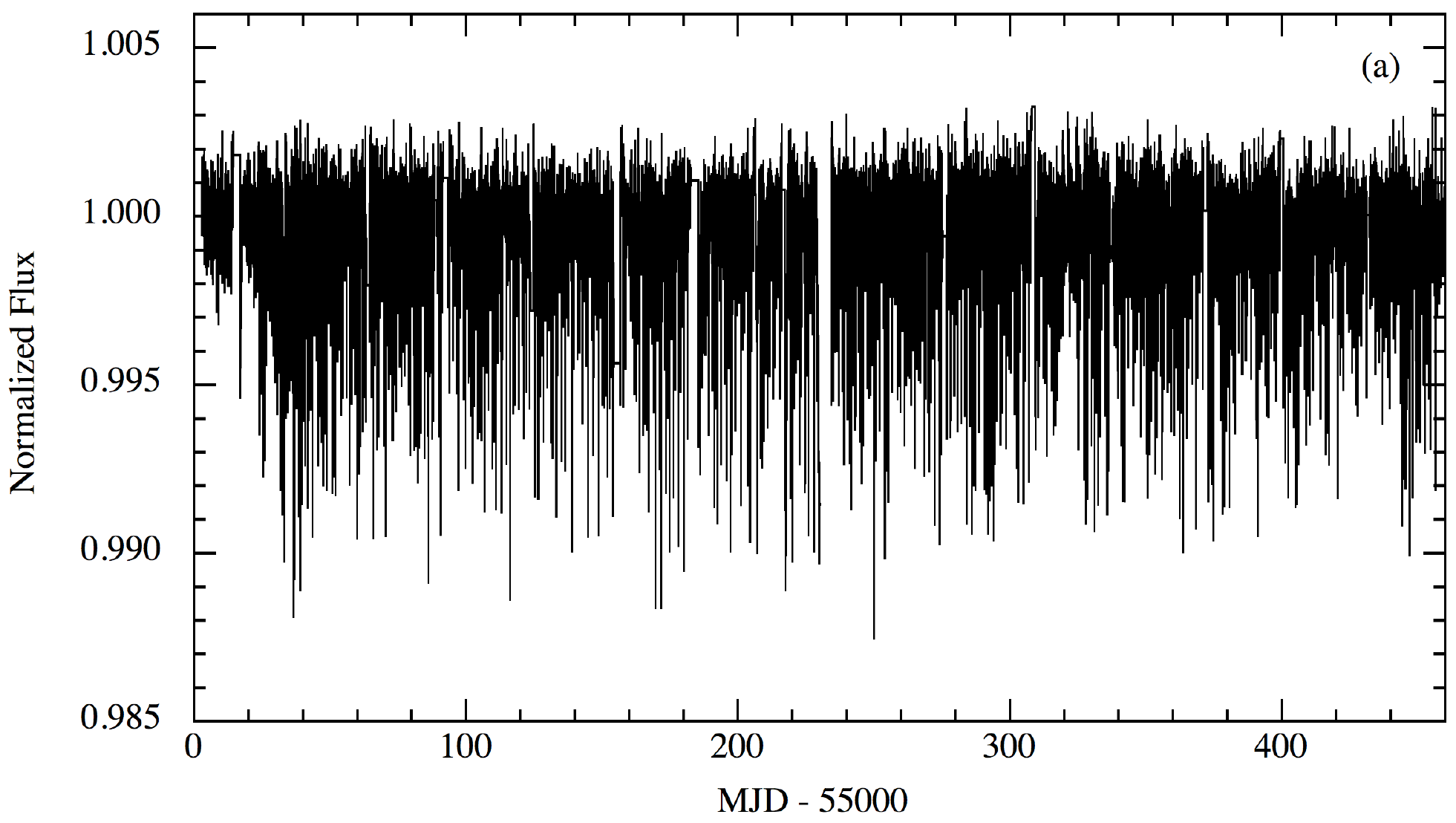} \vglue0.3cm
\hglue0.6cm \includegraphics[width=0.451 \textwidth]{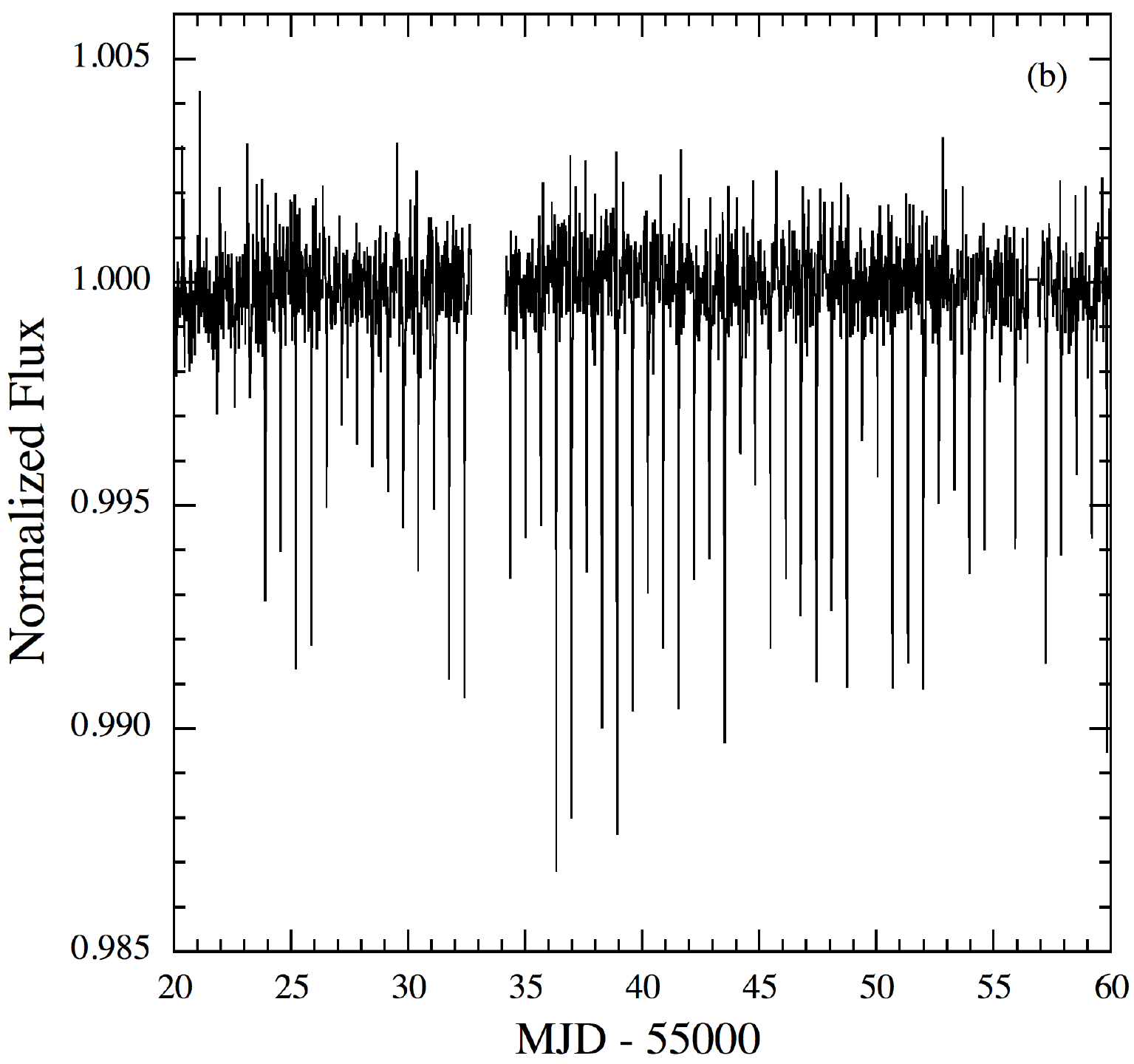} \hglue0.3cm
\includegraphics[width=0.430 \textwidth]{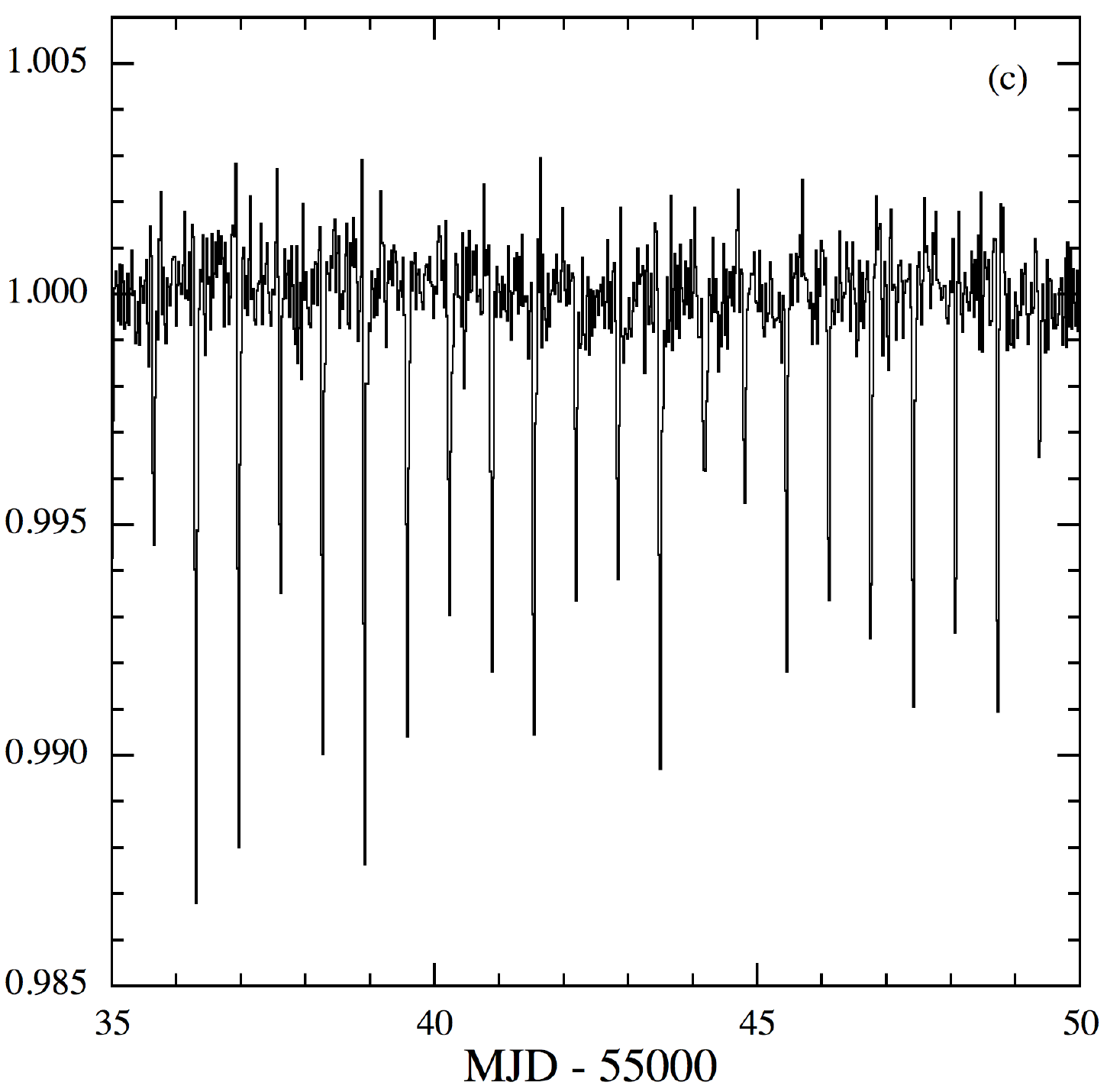}
\caption{Light curves for KIC 12557548 spanning the Q2--Q6 data sets, as well as for 40-day and 15-day segments of the Q2 data.  The intrinsic Poisson fluctuations per $\sim$30-min {\em Kepler} sample are $\sim$0.03\%.}
\label{fig:LCs}
\end{center}
\end{figure*}

One of the interesting power spectra contains a peak at 15.6854 hours
   and 15 harmonics thereof with amplitudes that decrease slowly with
   frequency.  This is indicative of a sharp
   transit/eclipse/occultation which occupies only a small fraction of
   a cycle.  The target, KIC 12557548, does not appear in either the
   Prsa et al.\,(2011) binary star catalog or the Borucki et
   al.~(2011) planet catalog.

\subsection{Light Curves}
\label{sec:LC}

Light curves for KIC 12557548 are shown in Fig.\,\ref{fig:LCs}. The
plots shown in this figure were made by combining the raw
(SAP$\!\_$FLUX) long-cadence light curves from quarters 2 through 6 (Q2--Q6);
before this was done the light curves from the various different
quarters were multiplied by relative scale factors chosen to minimize
any differences in intensity across quarter boundaries.  Low frequency
variations caused by instrumental, and possibly intrinsic source,
variations were then removed by convolving the flux data with a boxcar
of duration 0.65356 days (15.5854 hours) and then subtracting the
convolved flux curve from the pre-convolution light curve. This
processing should largely preserve the actual source behavior on
orbital and suborbital timescales.\footnote{We also used the PyKE
  software to process the data. See
  \url{http://keplergo.arc.nasa.gov/ContributedSoftwarePyKEP.shtml}}

Fig.\,\ref{fig:LCs} shows the highly variable nature of the
   occultations which, in fact, have depths that render them
   easily seen over most of the 460 day interval, but for the first
   $\sim$22 days of Q2 are very small or even not noticeable.  We
   have also checked the Q1 data and, indeed, the occultations
   are not apparent during the first 10 days as well as the final 10 days
   of that quarter.  As seen in panel (c) of Fig.\,\ref{fig:LCs}, the
   individual occultations vary strongly from one cycle to the next.
   The variable occultations persist, with a similar range in depths,
   throughout the following 438 days of Q2 through Q6 data.
   
\begin{figure}
\begin{center}
\vglue1cm
\includegraphics[width=0.98 \columnwidth]{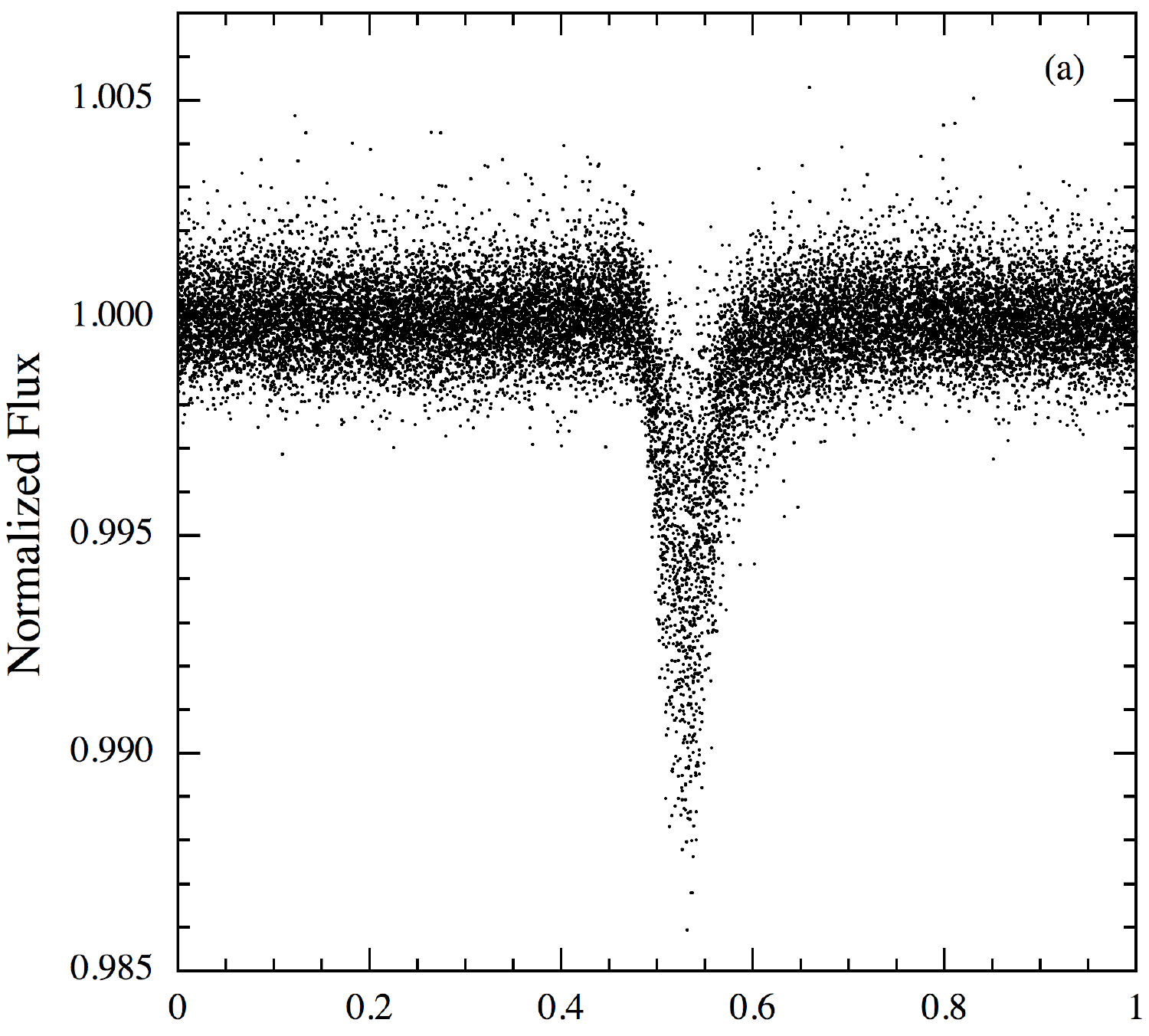}
\includegraphics[width=0.98 \columnwidth]{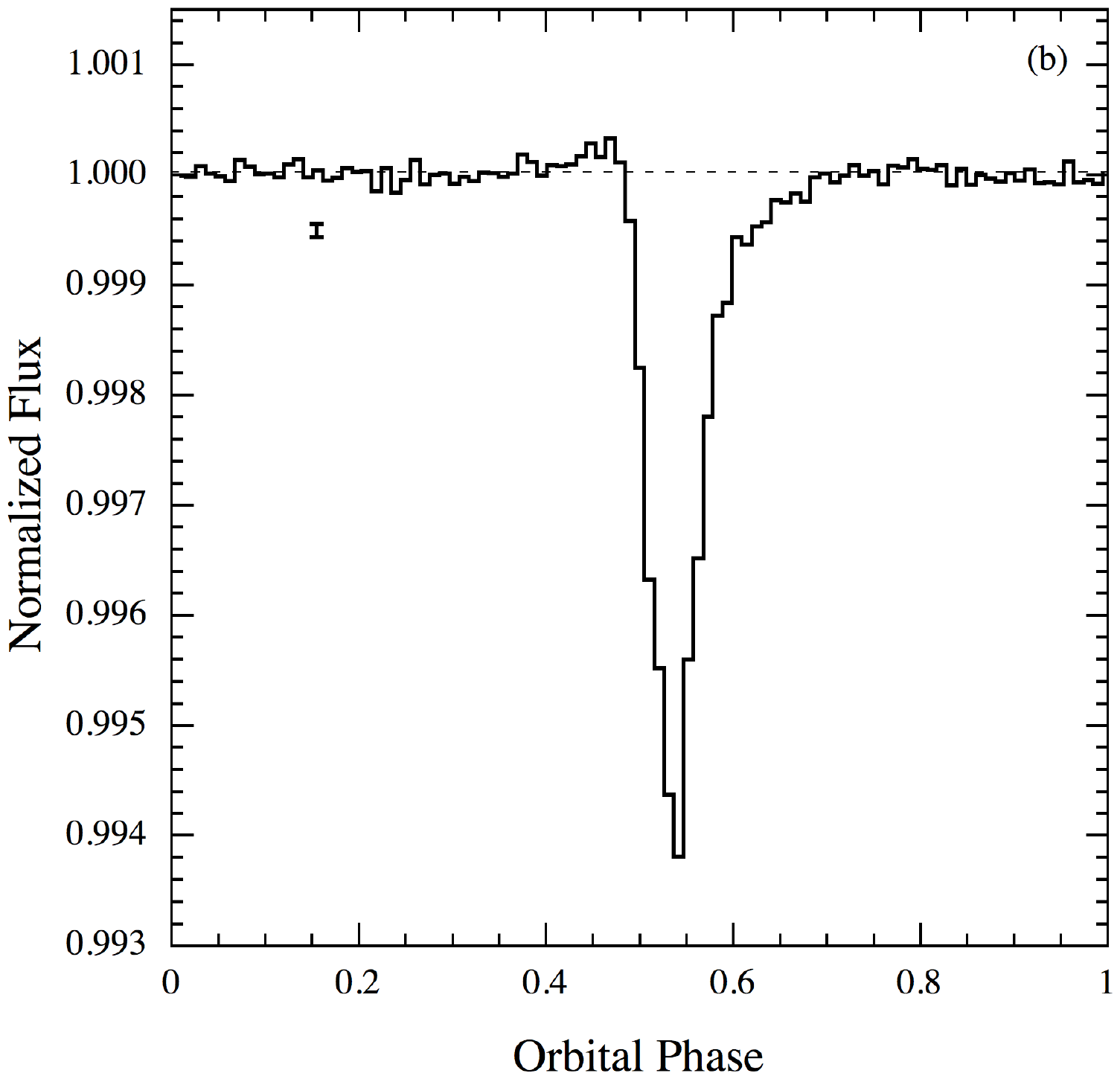}
\caption{Folded light curves of KIC 12557548 about the 15.685-hour occultation period for the Q2--Q6 data sets.  Top panel -- unbinned data; bottom panel -- folded data averaged into 96 discrete bins (3 bins $\simeq$ 1 long-cadence {\em Kepler} integration time).  Short illustrative vertical bar on the left is $\pm$ the standard error of the data points within a bin.  Note the highly statistically significant depressed flux level following the main occultation.}
\label{fig:folds}
\vspace{0.3cm}
\end{center}
\end{figure}

When the data are folded about the 15.685-hour period, the results,
   shown in Fig.\,\ref{fig:folds}, further illustrate how the
   occultation depths vary.  The largest depth is $\sim$1.3\%, while
   some of the points near the middle of the occultation are depressed
   by no more than $\sim$0.15\%.  On the other hand, it is clear that,
   aside from a few individual data points within the occultation
   interval, almost {\em all} show a minimum depression in the
   intensity by $\sim0.1$\%.  Thus, even when the
   occultations are not noticeable, there still may be a
   small effect.
   
When the folded data are binned
   (see Fig.\,\ref{fig:folds}b) one can see that the average depth at
   the heart of the occultation is only $\sim$0.6\% or about half
   the maximum depth.  There is evidence (at the $\sim$5-$\sigma$ level) 
   for a small peak in the flux of amplitude $\sim$$3 \times 10^{-4}$ 
   at phases up to $\sim$$30^\circ$ before the occultation ingress.
   There is a also a very significant depression or deficit in the expected
   flux over a similar range of phases just after
   egress.  We will discuss these features briefly in Section \S
   \ref{sec:scatt}.  

The full width of the occultation, with the exception of these
   small features, is 0.1 of the orbital cycle.  This corresponds to
   $\sim$1.5 hours in duration, or just 3 {\em Kepler} long-cadence
   integration times.  If this duration is interpreted simply as indicating the sum of the
   radii of the occulting ``bodies'' it corresponds to ($R_1+R_2)/a \simeq
   0.3$.  However, if we take into account approximately the effect of
   the finite integration time, then ($R_1+R_2)/a \simeq 0.24$.  Note
   that these estimates assume, without justification, equatorial as opposed to grazing occultations.

We have fitted a constant plus a cosine with a period of 15.685 hr and 
a cosine of half that period 
   to the folded light curve, excluding points inside the occultation interval
   (i.e., at orbital phases of $\pm 0.085$ cycles around mid
   occultation), in order to search for possible ellipsoidal light variations.  We 
   find only an upper limit of $\sim$$5 \times
   10^{-5}$ for the amplitude of such features.  This limit will be
   discussed below in the context of setting a constraint on the mass
   of any body orbiting the target K star.

   Finally, we find the epoch of mid-occultation in order to project
   ahead to future observations:
   \begin{eqnarray}
   t_n = 55399.824(2) + n \times 0.65356(1) ~{\rm MJD}~,
  \end{eqnarray}
  where the numbers in parentheses are the respective uncertainties in the last significant figure given.

\begin{figure}[t]
\begin{center}
\includegraphics[width=0.98 \columnwidth]{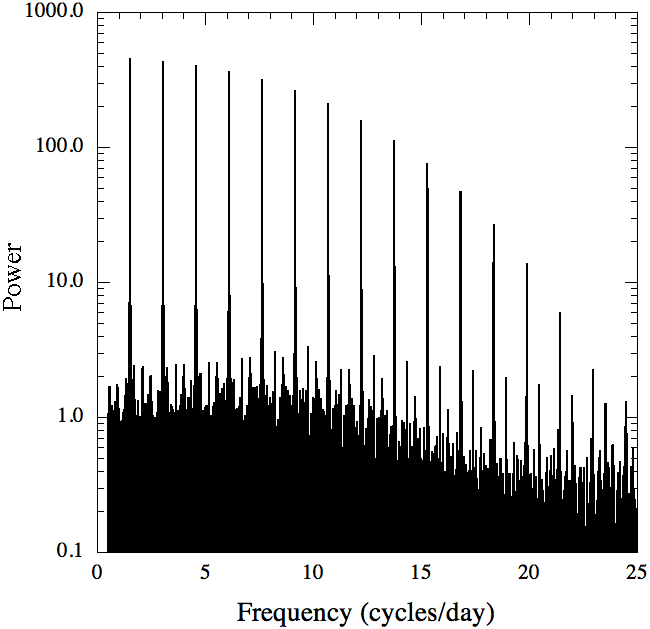}
\caption{Power spectrum of the detrended flux data for KIC 12557548 with the out-of-occultation region set equal to the mean flux level. See text for details.}
\label{fig:FFT}
\end{center}
\end{figure}

\subsection{Fourier Search for Modulations}

As a check for periodic modulations of the occultation depths for
   the 15.685-hour period, we carried out an FFT of the Q2 plus Q3 data 
    but with the portions of the light curve
   away from the occultation set equal to the mean out-of-occultation
   intensity.  The purpose of this latter step is to suppress the
   noise level, without sacrificing any significant fraction of the
   signal.  The results are shown in Fig.\,\ref{fig:FFT}.  All 16
   harmonics of the 15.865-hour period, out to the Nyquist limit, are
   clearly visible.  In addition, a careful inspection of the
   amplitudes between the harmonics indicates some evidence for
   low-amplitude modulation-induced sidebands.  However, at 
   least a number of these can be reproduced in an FFT of the 
   window function associated with the occultations.  Thus, we find no
   compelling evidence for periodic modulation of the occultation
   depths.

\subsection{Checks on the Validity of the Data}
\label{sec:valid}

Because of the unusual exoplanet phenomenon presented in this work, we need to be especially careful to ensure that there are no spurious artifacts in the {\em Kepler} data for this object. In this regard, we performed a number of tests on the data.

First, as mentioned above, we checked that the occultations are present in all of the quarters of released (i.e., public) data (Q1--Q6), and that the behavior of the occultations does not change abruptly across quarterly boundaries.

Second, we investigated whether the photometric variations of KIC
12557548 could be from another star or source on the sky. We examined
the Digitized Sky Survey images and found that there are no especially
bright stars close to KIC 12557548 whose light or transferred charge
might introduce spurious signals into the data stream (i.e., none with
V $<$ 13 within $3'$ and none with V $\lesssim$ 10 within $8'$). An
examination of a {\em Kepler} Full Frame Image for Q1 indicates that
there are no obvious stars within $\sim$$38''$ ($\sim$10 pixels). The
Kepler Input Catalog does list two $Kp$=19 stars located $14''$ and
$17''$ from KIC 12557548 (approximately 3.5 and 4.25 pixels), but
these could not be the source of the photometric variations for two
reasons. (i) We checked difference images for each of Q1--Q6 and the
apparent source of the photometric variations is coincident with the
stellar image of KIC 12557548 (Batalha et al.~2010).  Difference
images compare the in-transit frames with out-of-transit frames in the
neighborhood of each transit and these are averaged across all events
in each quarter. (ii) Analysis of the correlation of the astrometry
(brightness weighted centroids) with the photometric signature of the
occultations indicates that the position of KIC 12557548 shifts by no
more than $\sim$0.2 millipixels in either row or column during any
quarter.  
This indicates that there is no other background source that
is producing the occultations and that is offset by $\sim$2$''$ or
more from KIC 12557548, given the depth of the events and the brightness
of KIC 12557548 (Jenkins et al.~2010).
   
As a further check against video crosstalk from the adjacent CCD
   readout channels, we examined areas on the sky located at the same
   row/column position on these adjacent channels, and found no stars
   within $11''$ of the row/column position of KIC 12557548. The fact
   that the video crosstalk coefficients vary by typically 50\% or
   more from quarter to quarter as the stars are rotated about the
   focal plane, and can be either negative or positive, and
   the fact that the photometric signatures of the occultations are
   consistent from quarter to quarter, provide additional confidence
   that it is highly unlikely that the photometric variations are due
   to a source on an adjacent CCD readout channel.
   
Third, there are no binaries in the Prsa et al.~(2011) catalog or planets in the Borucki et al.~(2001) catalog with periods that match that of KIC 12557548.

\subsection{Optical Spectrum}
\label{sec:OMM}

\begin{figure}[t]
\begin{center}
\includegraphics[width=0.98 \columnwidth]{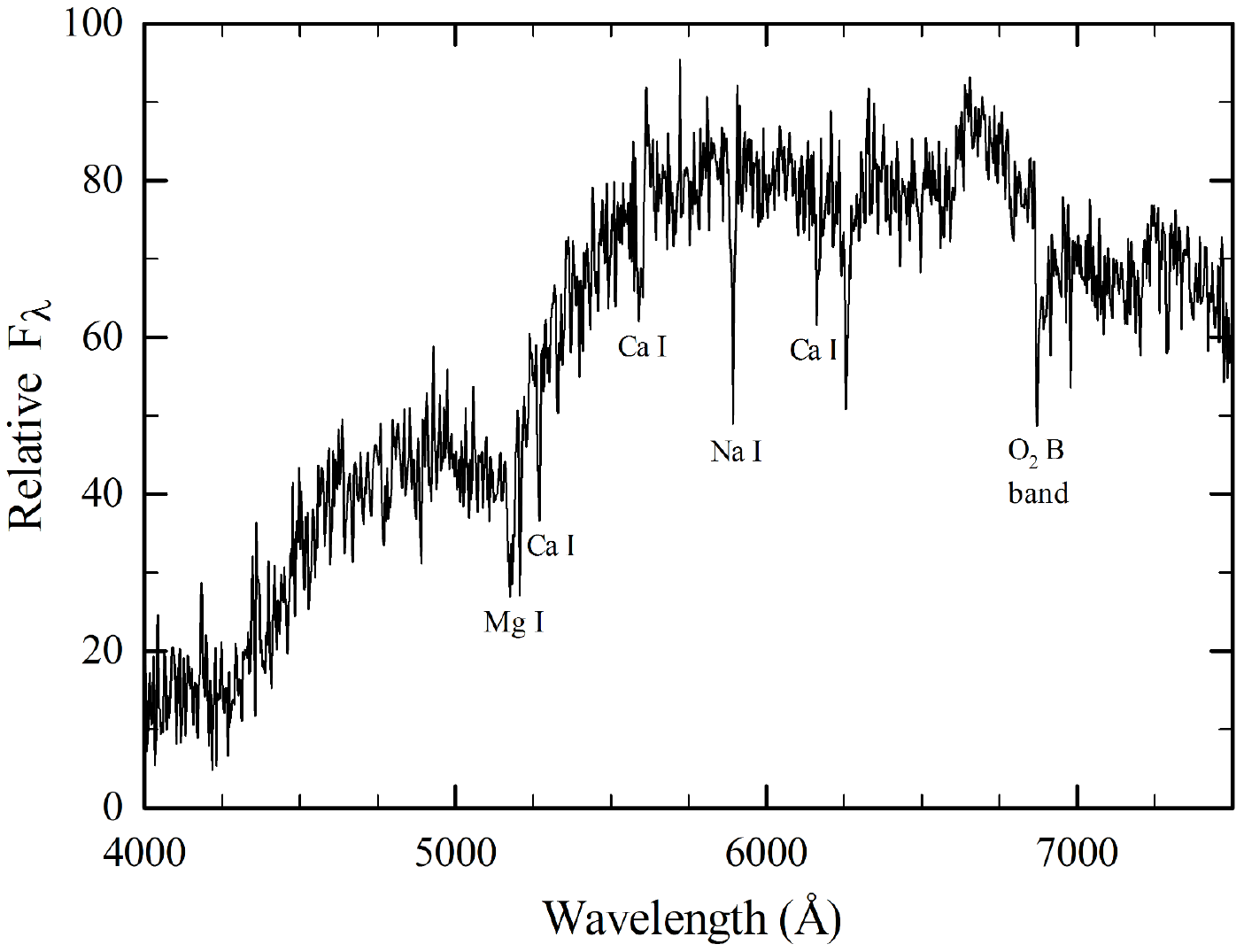}
\caption{Optical spectrum of KIC 12557548 taken on December 16, 2011 at the OMM Observatory.  The spectrum is a stack of four 15-minute exposures through $\sim$1.6 air masses.  The spectral features are indicative of a mid-KV star.}
\label{fig:OMM}
\end{center}
\end{figure}

KIC 12557548 is a $\sim$16th magnitude star located at RA(J2000) $=19^h~23^m~51.89^s$ and Dec(J2000) $= 51^\circ~30'~17.0''$.  We were able to obtain an optical spectrum of it on 2011 December 16 using the
   Observatoire Astronomique du Mont-M\'egantic's 1.6-m telescope.  Four 15-minute exposures were taken through 1.4 to 1.7 air masses soon after astronomical twilight. The spectrometer
   utilized the STA0520 blue CCD and a 600-line/mm grating, yielding
   an effective spectral resolution of 4.3 \AA.
   
The four spectra were background subtracted, corrected for the quantum efficiency of the detector and the reflectance of the grating, corrected for Rayleigh scattering in the atmosphere, and co-added to produce the spectrum shown in Fig.\,\ref{fig:OMM}.  We did not carry out absolute flux calibrations since the purpose of this observation was simply to confirm the spectral class
 of the {\em Kepler} target and to check for any unexpected spectral features.

Despite the relatively low signal to noise ratio of this spectrum, taken under non-ideal conditions, we can clearly see the signatures of a mid-K star. Comparisons with standard stars (see, e.g., Jacoby, Hunter, \& Christian 1984) indicate a spectral type of $\sim$K4V.  Furthermore, we find that the equivalent width of the Na I line is $3.6 \pm 0.4$ \AA \footnote{This is actually an upper limit to the EW because the ISM is expected to make a small contribution.} which is indicative of a K5-K7 star (Jaschek \& Jaschek 1987).  From this range of spectral types we infer $T_{\rm eff,s} \simeq 4300 \pm 250$ K, a mass of $M_s \simeq 0.70^{+0.08}_{-0.04} \,M_\odot$, and a radius of $R_s \simeq 0.65^{+0.05}_{-0.04} \, R_\odot$, assuming that the star is close to the ZAMS. These are all quite consistent with the values listed in the KIC. 

The KIC lists magnitudes of g=16.7, r=15.6, i=15.3, and z= 15.1 for
   star 12557548. From this we estimate the V magnitude is 16.2 (Windhorst
   et al.\,1991). If we take the absolute visual magnitude of the K
   star to be +7.6 (with $L_s \simeq 0.14\, L_\odot$), and correct for
   an estimated $A_V \simeq 0.22$, this puts the object at a distance
   of about 470 pc.

\section{Interpretation: General Remarks}
\label{sec:interpret}

The data analysis presented for KIC 12557548 in the previous
   sections reveals a highly periodic set of occultations, likely of
   the 16th magnitude K-star target of this particular {\em Kepler}
   field.  These occultations are highly variable in depth, and are
   therefore very different from anything yet reported about the other 
   couple of thousand transiting exoplanets (including candidates). 

   The remarkable stability of the occultation period strongly
   suggests that the occultations are due to an orbital companion to
   KIC 12557548. In \S\ref{sec:constrain}, we constrain the
   separation and mass of such a companion. These constraints are
   quite independent of any particular model. We then outline more
   specific scenarios to explain the variable nature of the
   occultations. These are: (1) direct occultations by one planet,
   modulated by a second planet that causes a precession of
   the orbital plane of the occulting planet (\S\ref{sec:dual}), and
   (2) dust that emanates from a single disintegrating planet
   (\S\ref{sec:disintegrate}). We consider both ideas qualitatively
   here, and then in \S \ref{sec:dust} we elaborate on the
   more likely one: that the occultations are due to dust emitted
   directly from a planetary atmosphere.  In \S
     \ref{sec:triple}, we discuss the alternative possibility that
     the occultations are not of the K star, but are rather due to a low-mass eclipsing stellar binary,
     possibly containing an accretion disk.

\subsection{Constraints on Separation and Mass \\ of Orbital Companion}\label{sec:constrain}

The regions of the light curve folded at the 15.685-hour period that
are away from the occultations show no evidence---with a limiting
amplitude of $\sim$5 parts in $10^5$---for ellipsoidal light variations 
(ELVs) with periodicities of 15.685 hr or half that. Since the amplitudes 
of ELVs are of order ${\rm ELV} \simeq (R_s/a)^3 (M_p/M_s)$
(where the subscript $s$ refers to the parent star) we can set a limit
on the mass of the companion $M_p$ (assumed $< M_s$)
if we have an estimate of the companion's orbital semimajor axis $a$.
Using the parameters for
KIC 12557548 taken from the KIC, as well as our
spectrum (see \S \ref{sec:OMM}), we find $T_{\rm eff,s} \simeq 4400$ K,
$\log g \simeq 4.63$, and $R_s \simeq 0.65 \,R_\odot$. 
These properties all point strongly toward a mid-K star of mass $M_s \simeq
0.7 \, M_\odot$.  The semimajor axis of a small body in an assumed
$P_{\rm orb} = 15.7$-hour orbit about such a star is
\begin{eqnarray}
a \simeq 2.8 \left(\frac{M_{\rm s}}{0.7\,M_\odot}\right)^{1/3} R_\odot \simeq 0.013 \, {\rm AU} \,.
\end{eqnarray}

We then have a rather good estimate of $R_{\rm s}/a$ of 0.23.
   When this value and our upper limit on the ELV amplitude are used 
   in the above expression for the ELV amplitude, we obtain a
   constraint on the mass of the orbiting companion of
   $$M_p \lesssim 3 ~M_J \,.$$ Thus, it is unlikely that the parent
   star has another star, or even a brown dwarf, orbiting it at a
   distance of only 2.8 $R_\odot$.\footnote{For an assumed orbit with
   $P = 2 \times 15.685$ hr, and two essentially equal eclipses per orbit, the 
   limit on the mass of the companion would be $\sim$11 $M_J$ (for $M_p \ll M_s$).}  
   This leaves only planetary-mass
   companions as the direct or indirect cause of the occultations.
      
\subsection{The Dual Planet Hypothesis}\label{sec:dual}

From our analysis in \S\ref{sec:constrain}, a planet similar to
Jupiter in mass and radius could be compatible with the lack of ELVs.
Such a giant planet would also generate maximum eclipse depths of
$\sim$1\%, like those observed. The challenge lies in having such a
Jovian-mass planet produce highly variable transit depths.

We could try to make such a model work by supposing that the
   transit is grazing, and that precession of the orbital plane alters
   the transit geometry so that the eclipse depth varies.  However,
   orbital precession induced either by tidal distortions or by
   another planet in an independent orbit generally occurs over
   timescales much longer than the orbital period. By contrast, the
   eclipse depths of KIC 12557548 vary markedly from orbit to orbit.
   Moreover, invoking a second planet in an attempt to force rapid
   orbit-to-orbit variations in the planet's trajectory across the star would
   result in transit timing variations (TTVs; Agol et al.~2005; Holman
   \& Murray 2005).  No TTVs are observed, although the long-cadence
   {\it Kepler} integration time hinders us from detecting TTVs
   substantially shorter than $\sim$30 minutes. In addition, the peculiar
   ingress-egress asymmetry---in particular the pre-ingress
   brightening---evinced in the {\it Kepler} light curve (Figure
   \ref{fig:folds}) finds no ready explanation under the dual planet
   hypothesis.

Another way to induce precession would be to have a second planet
   orbit the first in a binary planet configuration (Podsialowski et
   al.~2010). However, we have verified, using the dynamical stability
   criteria of Mikkola (2008, and references therein), that for the
   parameters of KIC 12557548, two Jupiter-sized planets cannot be in
   a stable mutual orbit and avoid Roche lobe overflow.

\subsection{Occultation by Debris from a \\ Disintegrating Rocky Planet}\label{sec:disintegrate}

Here we consider another scenario---that the occultations are
produced by a much smaller, rocky planet, heated to such
high temperatures that it is vaporizing. In evaporating away, it
produces a debris field of dust that variably obscures up to $\sim$1\%
of the light from the parent K star.

A purely gaseous wind emitted by a planet will not have sufficient
broadband opacity to affect significantly the transit depth in the
{\it Kepler} bandpass.  For example, the hot Jupiter HD 209458b is
known to emit a gaseous wind (Vidal-Madjar et al.~2003; Ben-Jaffel
2008; Linsky et al.~2010).  Although this wind increases substantially
the optical depth in narrow atomic absorption lines, it hardly alters
the transit depth in the broadband optical. For KIC 12557548, we must
consider instead that the planetary outflow contains particulates, e.g., dust.
Such dust would be driven off a rocky planet so hot that it is vaporizing.
This scenario is further developed and quantified in \S\ref{sec:dust}.

\subsection{Other Orbital Configurations} \label{sec:triple} 

  Could the source of the occultations be a low-mass eclipsing
  stellar binary that either is coincidentally located close on the
  sky to KIC 12557548 (but see the limits we have placed on blends in
  \S\ref{sec:valid}) or orbits KIC 12557548 in a hierarchical triple
  system?  Such a binary might consist of, e.g., a pair of M stars,
  whose strongly modulated light would be diluted by light from the K
  star seen in KIC 12557548. If such a binary had an orbital period
  about the K star of $\gtrsim$10 days, it could avoid producing a
  detectable ELV in the light curve of the parent star. But we know
  of no mechanism by which an eclipsing stellar binary like this,
  whether it is or is not in a hierarchical triple, can generate
  variable transit depths (e.g., orbital precession is too slow), let
  alone reproduce the transit profile shown in Figure 3 with its
  distinctive pre-ingress brightening and prolonged
  ingress.

Another possibility is that one component of the
  low-mass eclipsing stellar binary is a compact object, e.g., a
  white dwarf, fed by an accretion disk (G. Marcy, personal
  communication). Light from the binary, made variable by
  occultations of the accretion disk by the donor star, would be
  diluted by light from the K star. To explain the variable transit
  depths, the accretion disk itself would have to vary in
  luminosity. Why such luminosity variations would be restricted to
  those phases when the disk is occulted---i.e., why they would not
  also occur during the $\sim$70\% of the orbital cycle when the
  light curve is essentially flat---is not explained under
  this hypothesis.

\section{Obscuration by Dust from a Disintegrating Super-Mercury}
\label{sec:dust}

In this section we examine in some detail how dust emanating from a
possibly rocky planet might result in variable occultations of the
parent star of up to 1.3\% in depth.  We discuss the size of the
required dust cloud, composition and survival requirements for the
dust particles, and required mass loss rates and implied lifetime
of the planet against evaporation (\S\S \ref{sec:superearth},
\ref{sec:dustcomp}, \ref{sec:taumdot}).  A mechanism of ejecting gas
and dust from a planet with non-negligible gravity is described in \S
\ref{sec:eject}.  We argue that dust can provide a natural explanation
for the time variability of the mass loss and thereby of the variable
occultation depths (\S \ref{sec:limit}).  We carry out a numerical simulation of
the motions of dust particles after they escape from the planet's gravitational
well that shows that 
a comet-like tail is formed, and that illustrates how the dust might scatter
radiation as well as absorb it (\S\S \ref{sec:dustflow},
\ref{sec:scatt}).  We go on to discuss how long such a close planetary
system could last under the influence of tidal drag on the orbit (\S
\ref{sec:tidal}).  Finally, we explain why occultations by dust have
not been heretofore seen in other close-in, hot, rocky planets (\S
\ref{sec:difference}).

A number of these sections are necessarily speculative or involve
some assumptions.  We offer these simply to start the thinking
process on this subject rather than to provide definitive answers.

\subsection{The Underlying Planet vs. \\ the Size of the Dust Cloud}
\label{sec:superearth}

To set the stage, we assume in what follows that the planet underlying the
outflow (hereafter ``KIC 1255b'')
has an Earth-like composition, and has mass $M_p = 0.1 \,
M_{\oplus}$ (1.8 times the mass of Mercury) and radius $R_p = 0.5 \,R_{\oplus}$.
In the absence of
an outflow, our nominal ``super-Mercury'' would produce a transit depth of at most
$(R_p/R_s)^2 \approx 5 \times 10^{-5}$, consistent with our lack of detections
of transits in portions of the light curve.  The planet's Hill sphere radius
(distance from the planet to the nearest Lagrange point) is
\begin{eqnarray}
R_{\rm Hill}  =  \left( \frac{M_p}{3M_s} \right)^{1/3} a \,
  \simeq \,  1 \times 10^9 \left( \frac{M_p}{0.1 \,M_{\oplus}} \right)^{1/3} {\rm cm} \,.
\end{eqnarray}
The Hill sphere radius is smaller than $R_o$, defined as the
characteristic size of the occulting region during the deepest
eclipses (i.e., the size of the dust cloud).  We adopt a nominal value
for $R_o = 0.1\, R_s \simeq 5 \times 10^9 \, {\rm cm}$, which is such
that if the occulting region were circular and optically thick, it
would generate a transit depth of 1\%, near the maximum observed.  From our
experiments (not shown) simulating the occultation profile, we
estimate that $R_o$ could be a few times larger than our nominal
choice if the occulting region were optically thin and/or the transit
were not equatorial.

We have chosen a Mercury-like planet for our fiducial case
because a low escape velocity allows us to explain how dust might be
ejected.  The physics of how grains can be launched successfully from
a super-Mercury is explained in \S\ref{sec:eject}. In contrast, the
difficulties of ejecting dust from hot rocky super-Earths are
discussed in \S\ref{sec:difference}.

\subsection{Dust Composition and Survival}
\label{sec:dustcomp}

Dayside planet temperatures peak at approximately
\begin{equation} \label{eqn:teff}
T_{\rm eff,p} \simeq T_{\rm eff,s}\sqrt{\frac{R_{\rm s}}{a}} \simeq
2100 \, {\rm K} \,,
 \end{equation}
 neglecting, among other effects, transport of heat from the
 substellar point by the atmosphere or by a magma ocean (see
 L\'eger et al.~2011 for an estimate of the location of the shore of the
 magma ocean for the hot super-Earth CoRoT-7b).  Such nominal
 temperatures are hot enough to vaporize silicates and a variety of
 solid oxides, and to form a high-Z atmosphere.
 
 This particular model requires that the atmosphere contain refractory
 grains which can survive for long enough to produce an observable
 occultation. We can think of at least two ways of
 loading the atmosphere with dust. The first is by condensation of
 dust grains from a metal-rich vapor into clouds akin to those in
brown dwarf or giant planet atmospheres; see, e.g., the modeling of
forsterite clouds in Cooper et al.~(2003) and Madhusudhan et
al.~(2011).  Condensation can occur where gas cools, at altitude or
near the day-night terminator (Schaefer \& Fegley 2009; Schaefer et
al.~2011; Castan \& Menou 2011).
Whether such clouds can contain enough condensed mass remains
to be seen---we will see in \S\ref{sec:taumdot} and \S\ref{sec:eject}
that the planetary wind must contain at least as much mass in gas as
in solids (or liquid droplets). 

A second way to inject dust into the atmosphere is through explosive
volcanism. Tidal locking of hot rocky planets may produce a pattern of
mantle convection in which volcanism is focused near the substellar
point, where a thermal wind (\S\ref{sec:eject}) is best able to
entrain volcanic ash (Gelman, Elkins-Tanton, \& Seager~2011; van
Summeren, Conrad, \& Gaidos~2011).  Weakening and thinning of the
planetary lithosphere by sublimation may render the surface
particularly vulnerable to volcanic activity. Eruptions beyond the
shore of the magma ocean may be supplemented by violent bursting of
bubbles at the magma ocean surface: on Earth, such eruptions can eject
magma spray up to $\sim$400 m/s (Taddeucci et al.~2012).  Io's plumes
are a Solar System analog for volcanic injection of dust into a
tenuous atmosphere; dust grain radii range from 0.03--0.12 $\mu$m
(Jessup \& Spencer 2012).

For completeness, we note that there might also be a third way
 to pepper the atmosphere with dust: as the still solid portions of
 the planetary surface sublime, the resulting vapor may entrain grains
 of more refractory composition.

The upper mantle of the Earth is composed overwhelmingly of pyroxene
([Mg,Fe]SiO$_3$) and olivine ([Mg,Fe]$_2$SiO$_4$). We might expect
whatever grains to be present in the atmosphere of an evaporating
sub-Earth to also consist of these minerals, at least in
part.\footnote{Another possibility involves grains of alumina
  (a.k.a. corundum; Al$_2$O$_3$), which seems likely if the planetary
  surface has been distilled to the point where it is composed
  predominantly of calcium-aluminum oxides (see Figure 5 of L\'eger et al.~2011).
}
The observations of
 KIC 12557548 demand that grains not sublimate before they can cover
 up to $\sim$1\% of the face of the star. This requirement can be met
 by micron-sized grains of pyroxene, but not of olivine. In a study of
 grain survival times in the tails of sun-grazing comets, Kimura et
 al.~(2002; see their Figure 4) calculated that the sublimation
 lifetime of an amorphous pyroxene grain having a radius $s \sim 0.2
 \,\mu$m is $t_{\rm sub} \sim 3 \times 10^4$ s at a distance of $7.5
 \, R_{\odot}$ from the Sun (equivalent to a distance of $2.8 \,
 R_{\odot}$ from the K star in KIC 12557548, neglecting the difference
 in the shape of the stellar spectrum heating the grain). Crystalline
 pyroxene survives for still longer, a consequence of its lower
 absorptivity at optical wavelengths and hence cooler temperature. By
 contrast, micron-sized grains of olivine ([Mg,Fe]$_2$SiO$_4$)
 vaporize in mere seconds.\footnote{We have been unable to identify a
   simple reason why the sublimation times of olivine grains are orders of
   magnitude shorter than those of pyroxene grains; see references in Kimura
   et al.~(2002) for the original laboratory data.
}

A sublimation lifetime of $t_{\rm sub} \gtrsim 3 \times 10^4$ s, as is
afforded by micron-sized pyroxene particles, is long enough for grains
to travel the length of the occulting region. This travel time is on
the order of
\begin{equation} \label{eqn:t_travel}
t_{\rm travel} \sim \frac{R_{\rm Hill}}{c_s} + \frac{R_o}{v} \sim 2 \times 10^4 \, {\rm s} 
\end{equation}
where the first term accounts for the time to travel
from the planetary surface to the Hill sphere,
and the second term accounts for the time to travel out to $R_o$. 
For the first leg of the journey, we have adopted a flow speed
equal to the sound speed $c_s$ in the planet's high-Z atmosphere:
\begin{equation}
c_{s} \simeq \sqrt{ \frac{kT}{\mu m_{\rm H}} } \simeq 0.7 \left( \frac{T}{2000 \, {\rm K}} \right)^{1/2} \left( \frac{30}{\mu} \right)^{1/2} \, {\rm km/s}
\end{equation}
where $k$ is Boltzmann's constant, $\mu$ is the mean molecular weight,
and $m_{\rm H}$ is the mass of hydrogen. The sound speed is
appropriate to use if the grains are carried off the planetary surface
by a thermal Parker-type wind, as discussed in \S\ref{sec:eject}.  For
the second leg of the journey, we adopt $v \sim 10$~km/s. This is a
characteristic grain velocity once the wind is rarefied enough that
dust decouples from gas, and is given by the dynamics 
dictated by gravity and radiation pressure (see \S\ref{sec:dustflow}).

\subsection{Mass Loss Rate and Evaporation Timescale}
\label{sec:taumdot}

The mass loss rate in dust grains may be estimated as
\begin{equation}
\dot{M}_d \simeq \Omega \rho_d v R_o^2
\end{equation}
where $\rho_d$ is the mass density in dust grains at a distance $R_o$
from the planet,\footnote{For this back-of-the-envelope estimate,
we neglect the difference between the planet center and the center of 
the dust cloud.} $v$ is the outflow velocity (on scales of $R_o$),
and $\Omega$ is the solid
angle subtended by the flow as measured from the planet center. We can
relate $\rho_d$ to the optical depth through the dust flow,
integrated along the line-of-sight path of length $\mathcal{R} \sim R_o$:
\begin{equation} \label{eqn:tau}
\tau \simeq \rho_d \mathcal{R} \kappa_d \sim \frac{\rho_d R_o}{\rho_{\rm b} s}
\end{equation}
where the dust opacity $\kappa_d \sim s^2 / (\rho_{\rm b} s^3) \sim 1 / (\rho_{\rm b} s)$ for grains of radius $s$ and bulk density $\rho_{\rm b} \sim 3 \, {\rm g~cm}^{-3}$. We can also relate $\tau$ to the eclipse depth $f$, assuming
$\tau \lesssim 1$ (we will discuss why the flow may be optically thin, or marginally so, in \S\ref{sec:limit}):
\begin{equation}
f \sim \frac{\tau R_o^2}{R_s^2} \sim 0.01 \left( \frac{\tau}{1} \right) \left( \frac{R_o/R_s}{0.1} \right)^2 \,.
\end{equation}
Combining the above three relations, we have
\begin{eqnarray}
\dot{M}_d & \sim & \frac{\Omega \rho_{\rm b} s f R_s^2 v}{R_o}  \simeq  2 \times 10^{11} \left( \frac{\Omega}{1} \right) \left( \frac{\rho_{\rm b}}{3 \, {\rm g/cm}^3} \right) \left( \frac{s}{0.2 \, \mu{\rm m}} \right)  \nonumber \\
& &  \times \left( \frac{f}{0.01} \right) \left( \frac{v}{10 \, {\rm km/s}} \right) \left( \frac{0.1}{R_o/R_s} \right)  \left( \frac{R_s}{0.65\, R_{\odot}} \right) \, {\rm g/s} \nonumber \\
& & \nonumber \\
  & \simeq & ~1 \, M_{\oplus} / {\rm Gyr} 
\label{eqn:mdot}
\end{eqnarray}
where $\Omega$ is normalized to 1 to account for the
likelihood that the area from which dust is launched may
primarily be in the substellar region.  If, as noted in
\S\ref{sec:superearth}, $\tau \lesssim 1$ and/or the transit is not
equatorial, then $R_o$ may be up to a few times larger than $0.1 R_s$,
and the above estimate of $\dot M_d$ would need to be reduced somewhat.

Gas (metal-rich vapor) adds to the total mass loss rate by a factor of
$\xi$.  If we assume that the mass loss mechanism principally produces
a gaseous outflow, and that drag accelerates the dust grains, then it
seems reasonable to assume that the gas density must be at least as
large as the dust density: $\xi \gtrsim 2$.  Otherwise the
backreaction drag on gas by dust would slow the gas down and prevent
it from escaping.  We will revisit the factor of $\xi$ more
quantitatively in \S\ref{sec:eject}; for now, we note that $\xi \sim
2$ falls squarely in the range of values inferred for cometary
outflows of dust mixed with gas (Delsemme 1982; Fernandez 2005).  Then
the evaporation timescale for our nominal $M_p = 0.1 \,M_{\oplus}$
planet is
\begin{equation} \label{eqn:evap}
t_{\rm evap} \sim \frac{M_p}{\mathcal{F} \xi \dot{M}_d} \sim 0.2 \left( \frac{0.3}{\mathcal{F}} \right) \left( \frac{2}{\xi} \right) \, {\rm Gyr}
\label{eqn:tevap}
\end{equation}
where $\mathcal{F}$ is the fraction of time the planet spends losing
mass at the maximum rate (set by $f = 0.01$). Our estimate of $t_{\rm
  evap}$ is nominally shorter than the likely age of the star, but not
by an implausibly large factor.

\subsection{Ejecting Dust Grains Via a Thermal Wind}
\label{sec:eject}

In this section, we explore the possibility that dust is ejected from KIC 1255b
in a manner that resembles dust ejection from
comets (for a pedagogical review of comets, see
Rauer 2007). Gas sublimates and then flows off a comet nucleus at essentially the
sound speed, which can be dozens of times greater than the escape
velocity from the nucleus. As gas atoms stream toward the
cometopause---the surface of pressure balance between cometary gas and
the Solar wind---they are photoionized by stellar ultraviolet
photons. At the cometopause, cometary ions are rapidly picked up by
the magnetic field of the Solar wind, and accelerated up to the Solar
wind speed of hundreds of km/s.  This explains the well-known
observation that the ionized gas tails of comets are swept
anti-Sunward, in the direction of the Solar wind.  Sufficiently
small dust particles, on the other hand, escape the nucleus by virtue
of the hydrodynamic drag exerted by the escaping gas.

An important difference between a comet and KIC 1255b is the much higher
surface gravity of our nominal super-Mercury. The surface escape velocity
\begin{equation}
v_{\rm esc,surf} \simeq 5 \left( \frac{M_p}{0.1 M_{\oplus}} \right)^{1/2} \left( \frac{0.5 R_{\oplus}}{R_p} \right)^{1/2} {\rm km~s}^{-1} 
\label{eqn:vesc}
\end{equation}
is $\sim$7 times higher than the local sound speed $c_s$.  At first
glance, then, it seems unlikely
that gas can be driven off the surface of the planet. But in 
Parker-type winds (see, e.g., Lamers \& Cassinelli 1999) the base of
the wind can be characterized by just such values of $v_{\rm
  esc,surf}/c_s$. Starting at their base, Parker-type winds accelerate
by gas pressure over long distances---several planetary radii---before
the bulk wind speed $v$ formally exceeds the {\it local} escape
velocity.  For
example, $v_{\rm esc,surf}/c_s \sim 5$ in the Solar corona, which is
the seat of the Solar wind (e.g., Lemaire 2011).  Another example is
provided by the thermal winds from hot Jupiters, where $v_{\rm
  esc,surf}/c_s \sim 5$ at the base of the planetary wind where
stellar UV photons are absorbed and heat the upper atmosphere (e.g.,
Murray-Clay, Chiang, \& Murray 2009).
Note that both these examples represent flows that are continuously
heated and energized along their extents---by photoionization in the
case of the hot Jupiter wind, and by conduction and wave heating in
the case of the Solar wind (e.g., Marsch, Axford, \& McKenzie 2003).
Thus these winds can remain nearly in the hydrodynamic regime and 
enjoy large mass fluxes, even at the large Jeans parameters
($\lambda_0 \equiv v_{\rm esc}^2/2c_s^2$) characterizing their base
radii $R_{\rm base}$.\footnote{
To quantify this statement, note that a hot Jupiter wind (as
computed, e.g., in the standard model of Murray-Clay et al. 2009) is
characterized by $\Phi/\Phi_{0,0} \approx 0.05$ at $\lambda_0  \approx
13.5$, in the notation of Volkov et al.~(2011). Here $\Phi/\Phi_{0,0}$
is the dimensionless mass loss rate, and the "0" base radius is taken
to be the surface which presents unit optical depth to UV photons. For
comparison, Volkov et al.~(2011), who neglect energy deposition at
radii $> R_{\rm base}$, predict $\Phi/\Phi_{0,0} \approx 4 \times
10^{-5}$ at $\lambda_0 = 13.5$ (see their Figure 3). Thus the
continuous energy deposition enjoyed by a hot Jupiter wind increases
the mass loss rate above what it would be without continuous energy
deposition by about three orders of magnitude.}

The thermal winds noted above suggest by analogy that
the atmosphere of KIC 1255b can be gradually accelerated by gas
pressure to speeds that eventually exceed the local escape velocity.
Gas can be continuously heated along the extent of the wind by
collisions with grains, which in turn are heated by starlight; thus
the wind should maintain a nearly constant temperature $\gtrsim 2000$
K.  At distances of $R_{\rm Hill} \sim 3 R_p$, near the first and
second Lagrange points where the effective gravitational acceleration
is nearly zero, we anticipate that the outflow will have attained the
sonic point, so that the wind velocity will be of order $c_s$ on these
scales (see, e.g., Figure 9 of Murray-Clay et al.~2009). This explains
the first term in equation \ref{eqn:t_travel}. Of course, this sketch
needs to be confirmed with a first-principles calculation of
the mass loss rate as a function of planet mass.

How dense and fast must the gaseous outflow be to lift grains off the
surface of the planet? The aerodynamic drag force on a grain must
exceed the force of planetary gravity:
\begin{equation} \label{eqn:liftoff}
\rho_g v_g c_s s^2 \gtrsim \rho_{\rm b} s^3 g
\end{equation}
where $g \sim 400$ cm s$^{-2}$ is the surface gravitational acceleration,
and $\rho_g$ and $v_g$ are the gas density and flow velocity, respectively.
For the drag force on the left-hand side we have used the
Epstein (free molecular) drag law, appropriate for grains whose sizes
are less than the collisional mean free path between gas molecules,
and for flow velocities $v_g \lesssim c_s$ (Epstein 1924).  Both of these
conditions can be verified {\it a posteriori} to hold throughout the
flow. 
The atmospheric density just above the ocean melt is $\rho_g \sim
P/c_s^2$, where the surface pressure $P \sim 10$ $\mu$bar at $T = 2100$ K, if the ocean is
composed predominantly of Mg, Si, and O, with zero contribution from
Na (see Figures 5 and 6 from L\'eger et al.~2011, and consider
vaporized fractions $0.85 > F_{\rm vap} > 0.02$; see also Schaefer \& Fegley 2009). 
Such an atmosphere is composed primarily of Mg, SiO, O, and O$_2$ gas, according 
to Figure 3 of Schaefer \& Fegley (2009). Then the vertical wind velocity
required to keep grains aloft against gravity is
\begin{equation} \label{eqn:breeze}
v_g \gtrsim 2 \left( \frac{10 \, \mu{\rm bar}}{P} \right) \left( \frac{s}{0.2 \, \mu{\rm m}} \right) \, {\rm m}/{\rm s}
\end{equation}
which seems easy to satisfy. If the surface has been vaporized
(distilled) to the point where it is composed of Al, Ca, and O
($F_{\rm vap} > 0.85$), then $P \sim 0.1$ $\mu$bar and $v_g \gtrsim
200$ m/s, which still seems possible (at least horizontal wind
velocities are thought to have near-sonic velocities; Castan \&
Menou 2011).

The lift-off condition (\ref{eqn:liftoff}) implies a mass loss
rate in {\it gas} of
\begin{equation} \label{eqn:mdot-gas}
\dot{M}_g \sim \Omega \rho_g v_g R_p^2 \gtrsim 4 \times 10^{10} \left( \frac{\Omega}{1} \right) \left( \frac{s}{0.2 \, \mu{\rm m}} \right) \, {\rm g/s} \,.
\end{equation}
The fact that the minimum value of $\dot M_g$ is comparable to $\dot{M}_d$ as estimated in
(\ref{eqn:mdot}) supports our claim in \S\ref{sec:taumdot} that the
gas-to-dust ratio may be of order unity, i.e., $\xi \sim 2$.

It is remarkable that at a maximum eclipse depth $f=0.01$, our
order-of-magnitude estimate for the total mass loss rate $\xi
\dot{M}_d \sim 4 \times 10^{11}$ g/s from our super-Mercury either
matches or exceeds, by up to an order of magnitude, estimates of mass
loss rates---mostly in hydrogen gas---from hot Jupiters like HD
209458b (Vidal-Madjar et al.~2003; Yelle 2004; Garcia-Munoz 2007;
Murray-Clay et al.~2009; Linsky et al.~2010) and HD 189733b
(Lecavelier des Etangs et al.~2010).  Hot Jupiters are thought to lose
mass by photoevaporation, whereby stellar X-ray and ultraviolet (XUV)
radiation photoionizes, heats, and drives a thermal wind off the
uppermost layers of a planetary atmosphere.  Present-day mass loss
rates of hot Jupiters are limited by the XUV luminosities of
main-sequence solar-type stars, of order $L_{\rm XUV,s} \sim
10^{-5}$--$10^{-6} L_{\odot}$.
In the case of KIC
1255b, mass loss may also be driven by a thermal
wind---but one which is powered by the bolometric stellar luminosity,
$L_s \sim 0.14 L_{\odot}$, which heats the entire planetary atmosphere
to temperatures $T \simeq 2000$ K.  The power required to
drive our inferred mass loss rate
\begin{eqnarray}
L_{\rm required} & = & \frac{G M_{\rm p} \xi \dot{M}_d}{R_{\rm p}} \nonumber \\
 & \sim & 6 \times 10^{22} \left( \frac{\xi}{2} \right) \left( \frac{s}{0.2 \, \mu{\rm m}} \right) \left( \frac{\tau}{1} \right) \, {\rm erg/s} \,, \nonumber \\
 & &
\label{eqn:Lreq}
\end{eqnarray}
($G$ is the gravitational constant) is easily supplied by the total power 
intercepted in stellar radiation,
\begin{equation}
L_{\rm intercepted} = \frac{L_{\rm s}}{4\pi a^2} \cdot \pi R_{\rm p}^2 \sim 4 \times 10^{26} {\rm erg/s} \,.
\label{eqn:Lint}
\end{equation}

\subsection{Time Variability of Mass Loss and \\ Limiting Mass Loss Rate}
\label{sec:limit}

The variability observed in the obscuration depths implies substantial
changes in $\dot{M}_d$ (or in other physical properties of the outflowing
dust clouds) over timescales $\lesssim P_{\rm orb} = 15.7$
hr.  Although we cannot be quantitative, such rapid variability seems
possible to accommodate within the framework of an evaporating planet.
Again, a comparison with comets is instructive. As in a cometary
nucleus, inhomogeneities or localized weaknesses in the planet's
surface (``vents''; Jewitt 1996; Hsieh et al.~2010) may cause the
planet to erupt occasionally in bursts of dust.  Whatever gas is
ejected from the planet will be photoionized by the stellar radiation
field and, like the gas tails of comets, interact with the magnetized
stellar wind. To the extent that dust can be entrained in plasma 
along the planet's magnetotail, plasma instabilities may introduce further 
time variability. 

The mass loss rate $\dot{M}_d$ may be limited to $\sim$$2 \times
10^{11}$ g s$^{-1}$, the value in equation (\ref{eqn:mdot})
appropriate to an optical depth $\tau \sim 1$ (assuming all other
parameters in that equation are fixed at their nominal values).  This
upper limit exists because if the flow became too optically thick,
light from the star would not reach the planet's surface to heat
it. Thus the mass loss rate may be self-limiting in a way that keeps
the outflow from becoming too optically thick.  One can imagine a
limit cycle which alternates between a low-obscuration phase during
which starlight strikes the planet's surface directly and dust
proceeds to load the planet's atmosphere, and a high-obscuration phase
during which dust removal dominates production.

Wind mass loss rates should depend on outgassing rates from the
planetary surface. Since saturation vapor pressures depend
exponentially on temperature, it seems easy to imagine that modest
changes in surface temperature due to changes in the optical
depth of the wind will drive large changes in $\dot{M}_d$.  As
noted above, a first-principles calculation of $\dot{M}_d$ along
the lines of the externally irradiated solution for hot Jupiter winds
(e.g., Murray-Clay et al.~2009) would be welcome.

\subsection{Dust Stream Structure}
\label{sec:dustflow}

\begin{figure}[t]
\begin{center}
\includegraphics[width=0.97 \columnwidth]{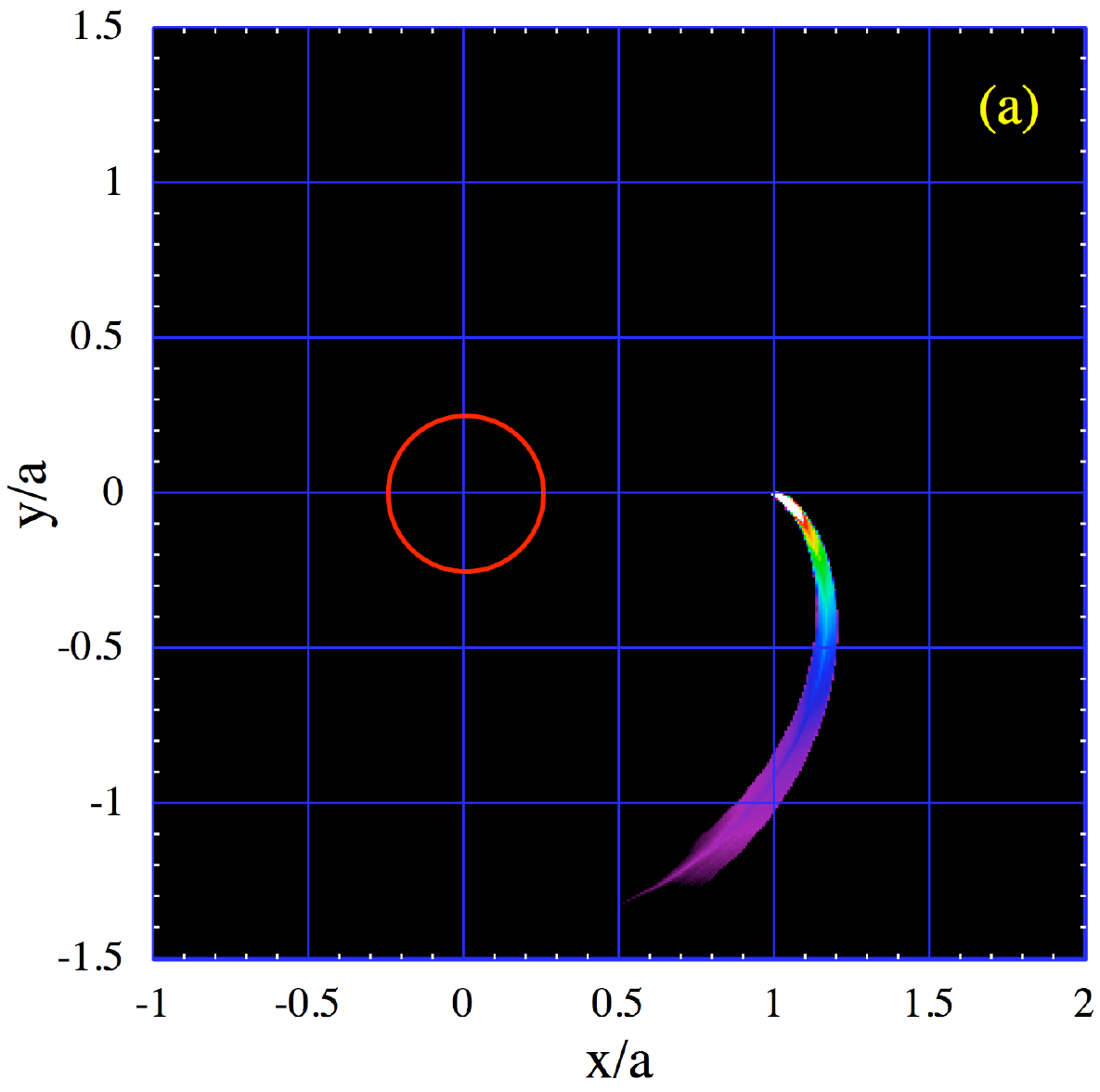}
\includegraphics[width=0.99 \columnwidth]{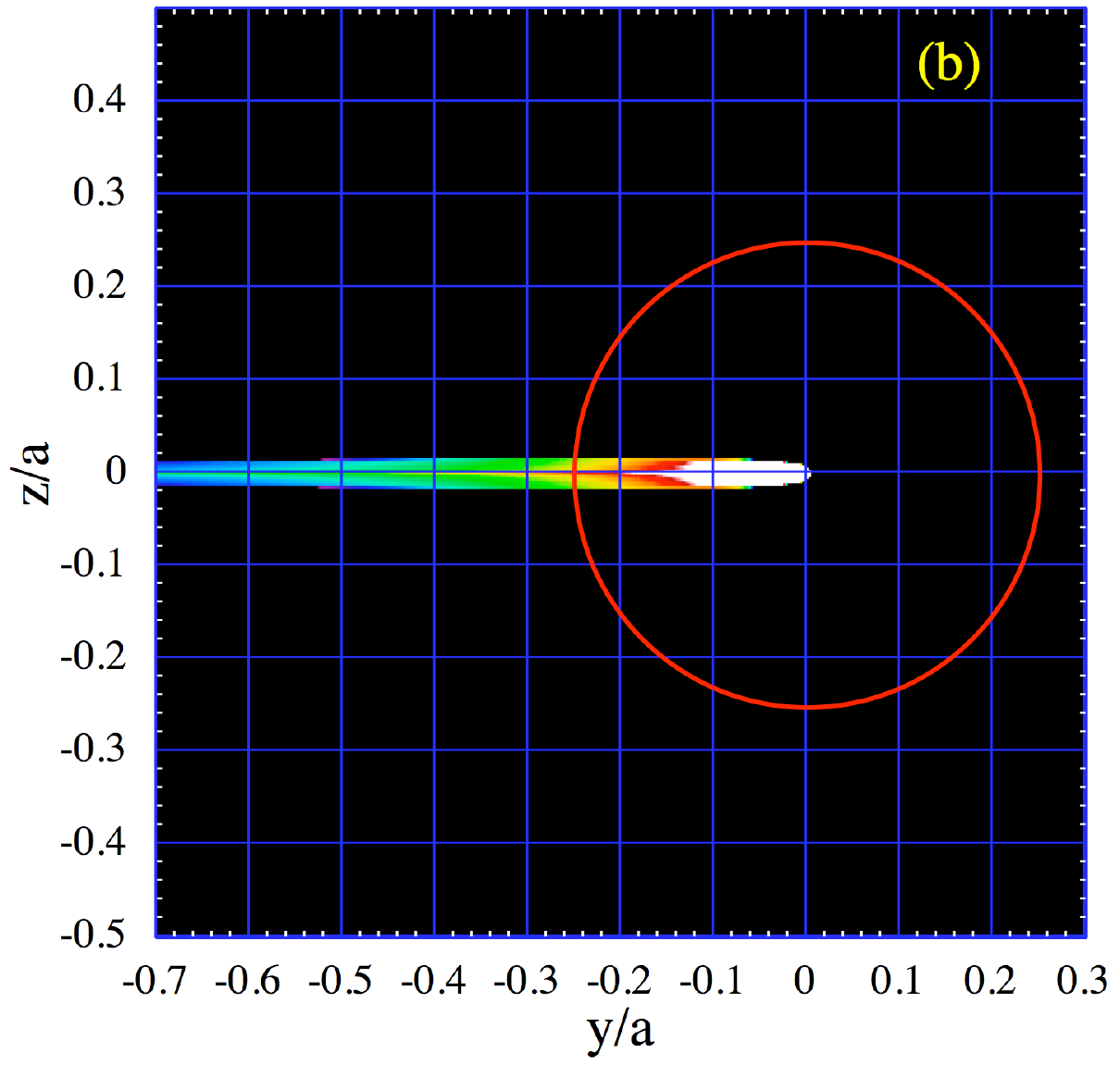}
\caption{Simulations of dust flows from a 0.1 $M_\oplus$ planet of 0.5 $R_\oplus$ driven by radiation pressure from the parent star.  The spatial scale is in units of the semimajor axis of the planet.  The color scale is proportional to the square root of the projected density in order to enhance the dynamic range.  The red circle represents the parent star. The top panel shows the dust
density as viewed from the orbital pole; the bottom
panel shows the dust density as seen by an
equatorial observer at mid transit.}
\label{fig:dust}
\end{center}
\end{figure}

Once dust escapes from the planet, and the gas density becomes too low
to exert significant drag, the dust will be swept back
into a comet-like tail by the combination of radiation pressure
and Coriolis forces. The ratio of the stellar radiation pressure
force on a grain to that of {\it stellar} gravity equals
\begin{eqnarray}
\beta & \equiv & \frac{F_{\rm rad}}{F_{\rm grav,s}}  =  \frac{3 Q_{\rm pr} L_s}{16\pi G c M_s \rho_{\rm b} s}  \nonumber \\
& \simeq & 0.15 \left( \frac{L_s}{0.14 \, L_{\odot}} \right) \left( \frac{0.7 \, M_{\odot}}{M_s} \right) \left( \frac{0.2 \, \mu {\rm m}}{s} \right)  \label{eqn:kimura} 
\end{eqnarray}
\noindent where $Q_{\rm pr}$ is the radiation pressure efficiency
factor (see, e.g., Burns et al.~1979) and $c$ is the speed
of light.  For the numerical evaluation in (\ref{eqn:kimura}), we have
used the results of Kimura et al.~(2002, their Figure 2) for a
spherical pyroxene particle, scaled for the luminosity and mass of our
K star.  The numerical result in (\ref{eqn:kimura}) is valid only for
$s \gtrsim \, 0.2 \,\mu$m (as $s$ decreases below 0.2 $\mu$m, $Q_{\rm pr}$
decreases and radiation pressure is increasingly negligible; in any
case such grains are unimportant because they sublimate too quickly).

We performed a simple calculation of what a dust tail might look
like.  Since the precise details of how dust is launched from
the surface are presently unclear---they are subject, e.g., to
horizontal winds on the planet that can distribute material
from the substellar point to the nightside---the calculation
simply assumes a spherically symmetric outflow of dust from the planet
at the surface escape speed.  Our fiducial planet parameters of $R_p =
0.5 \,R_\oplus$ and $M_p = 0.1 \,M_\oplus$, $\beta = 0.15$, and a
sublimation time for the dust of $t_{\rm sub} = 10^4$ s were 
assumed. The computations were performed in a frame of reference
rotating with the planet's orbit, and followed 30,000 dust particles
from launch to sublimation.

\begin{figure}[t]
\begin{center}
\includegraphics[width=0.98 \columnwidth]{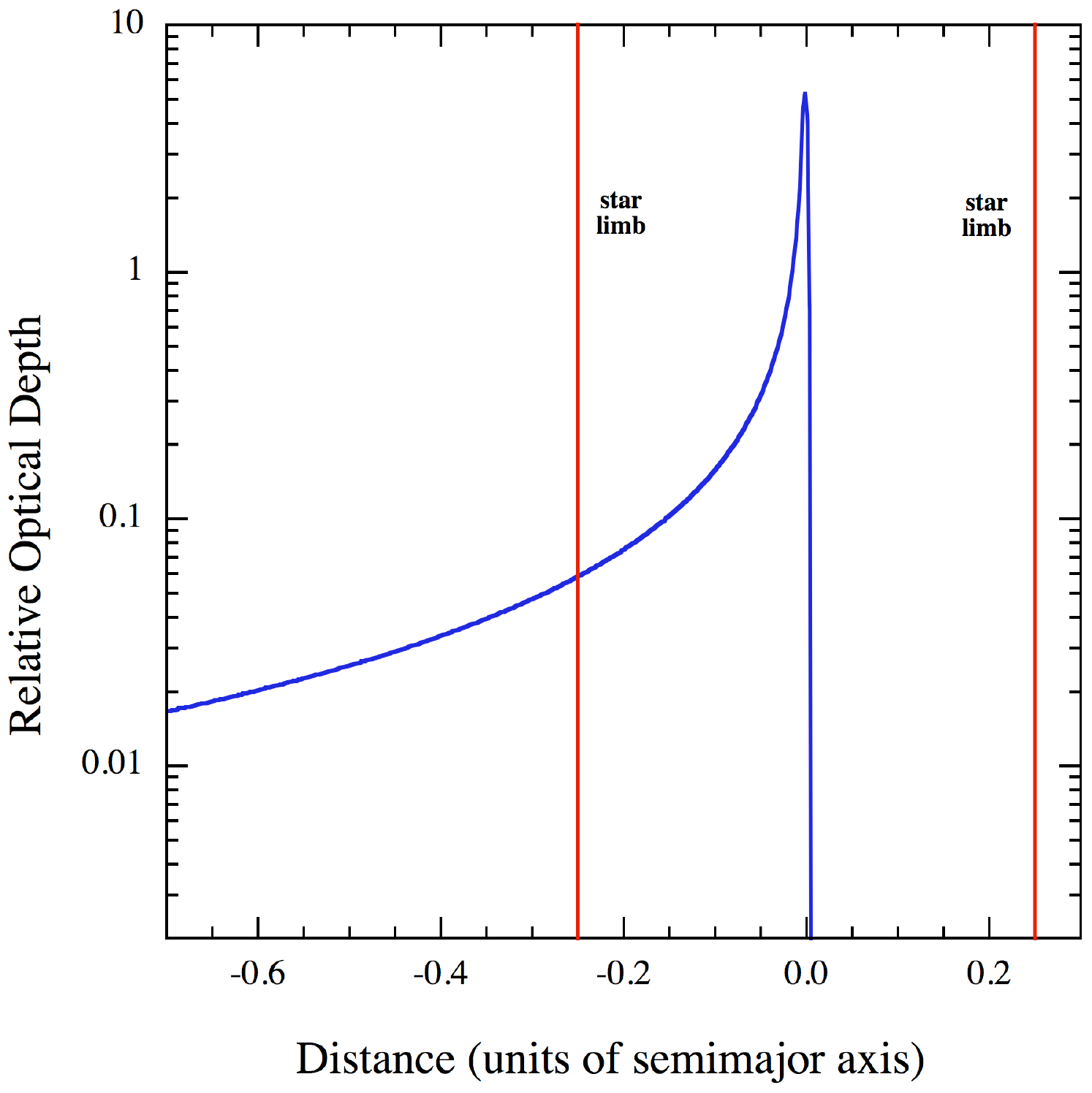}
\caption{Relative optical depth of the simulated dust tail as a
  function of distance across the parent star in units of the
  semimajor axis.  This result is derived from the simulated data used
  to produce Fig.\,\ref{fig:dust}.  The vertical scale is arbitrary
  since otherwise we would need to know the absolute value of the
  optical depth near the surface of the planet.}
\label{fig:profile}
\end{center}
\end{figure}

The results of this simulation are shown in
Fig.\,\ref{fig:dust}.  A long dust tail is apparent
because the color scaling covers a large dynamic range of column
densities.  The effective length of the tail is not actually that
long, e.g., the column density of dust particles is quite small at
distances comparable to $R_s$. 
This was quantified by integrating the total number of dust grains
within $\pm$ 1/8 of the stellar radius in height ($z$) as a function
of distance along the dust tail ($y$).  The result is shown in
Fig.\,\ref{fig:profile}.  Given the presently available information,
only relative values of the optical depth can be shown, but the curve
serves to show that the projected density of dust grains falls off
sharply with distance from the planet.  Thus, the optical depth could
be somewhat greater than unity near the projected planet disk, and the
net obscuration of the parent star could take on values up to the
maximum level of $\sim$1.3\% that is observed.

In summary, the dust stream geometry seems able to accommodate
the occultation durations and depths. It might even be able to
explain the curious ingress-egress asymmetry in the occultation profile,
as we argue in the next section.

\subsection{Scattering}
\label{sec:scatt}

The putative dust cloud likely attenuates more by scattering photons
out of our line of sight than by actual absorption (this is due to the
small imaginary part of the index of refraction in the visible; see,
e.g., Denevi et al.~2007).  This opens the possibility that
scattering of starlight into the line of sight would produce, at
certain orbital phases, an {\em excess} of light above the nominal
out-of-occultation region levels.  An excess would be more likely
given the presence of larger ($s \gtrsim 1 \,\mu$m) grains that
scatter primarily in the forward direction.

We can estimate the ratio of fluxes scattered into, versus out of, our line
of sight.  Consider an idealized, small, spherical scattering cloud of
radius $R_d$ that scatters a fraction $A$ (for ``albedo'') of all the stellar
radiation incident upon it (i.e., the cloud presents a total
scattering cross section of $A \cdot \pi R_d^2$). The incident
radiation is scattered uniformly into a solid angle $\Omega_{\rm scat}$ centered
about the direction of the incident beam.
In the case that the cloud sits directly in front of the target star,
the ratio of fluxes scattered into, versus out of, the
line of sight is
\begin{eqnarray}
\frac{\chi_{\rm in}}{\chi_{\rm out}} \simeq
\frac{\pi}{\Omega_{\rm scat}}\left(\frac{R_s}{a}\right)^2 \simeq
\frac{\pi}{16\,\Omega_{\rm scat}}
\end{eqnarray}
independent of $A$ and $R_d$. The ratio ranges from $\sim$1.6\% for
isotropic scattering, to as high as $\sim$20\% for forward scattering
into 1 steradian. If the dust cloud is blocking the stellar disk, then
attenuation is dominant (except in the case of extreme forward
scattering).  However, if the dust cloud is not blocking the stellar
disk, then there may be a small excess contribution to the system
light.  This may explain the small bump in flux just before ingress,
and the small depression in the flux just after egress (see
Fig.\,\ref{fig:folds}).  In this interpretation, the ingress-egress
asymmetry results from the asymmetric distribution of dust in the
planetary outflow.  Just prior to ingress, light is forward-scattered
off the dusty head of the ``comet'' into our line of sight, producing
the small bump in flux. Light may continue to be forward scattered
into the line of sight during the main occultation and even after
egress.  However, at those phases when part of the tail occults the
star, the scattering of starlight out of the line of sight is likely
to dominate and produce a net depression in the observed intensity.

\subsection{Tidal Inspiral Lifetime}
\label{sec:tidal}

We now consider how long such a planet as the one we are proposing
might last in the face of tidal decay due to interactions with the
parent star.  The timescale of tidal decay is approximately
\begin{eqnarray}
\tau_{\rm tidal} \simeq Q~ \frac{2 M_s}{9 M_p} \left(\frac{a}{R_{\rm s}}\right)^5 \frac{P_{\rm orb}}{2\pi} \simeq 10^{10} \left(\frac{Q}{10^6}\right)~{\rm yr}
\end{eqnarray}
(see, e.g., Gu, Lin, \& Bodenheimer 2003) where $Q$ is the usual
dimensionless measure of dissipation in the parent star, and we have
assumed that the parent star is rotating with an angular frequency
much lower than the planet's orbital frequency, and that the orbital
eccentricity is zero. Thus, even for a pessimistically small value of
$Q$, the lifetime against tidal decay is more than adequately long.

\subsection{Differences Between KIC 12557548 and Hot Super-Earths}
\label{sec:difference}

If our proposed scenario for KIC 1255b is correct, why don't other hot
rocky planets also evince occulting dust clouds?  The class of
close-in rocky planets includes CoRoT-7b (L\'eger et al.~2009; L\'eger et
al.~2011), 55 Cnc e (Winn et al.~2011), and Kepler-10b (Batalha et
al.~2011b).\footnote{Kepler-9d (Batalha et al.~2011a) may also belong
  to this class based on its size and orbital distance, but,
  unfortunately, its mass is not yet known, and therefore we cannot be
  sure that it is a purely rocky planet.}  All are argued to have
significant rock components, although an admixture of water, hydrogen,
and/or other light elements seems required to explain 55 Cnc e.

The primary parameter distinguishing the super-Mercury that we are imagining
orbits KIC 12557548 and the other hot rocky planets is clearly the planet
mass.  All the other planets are ``super-Earths'' having masses
$M_{\rm p} \simeq 10 \, M_{\oplus}$ and radii $R_{\rm p} \simeq 2 \,
R_{\oplus}$ (mass-radius relations for super-Earths, may be found in,
e.g., Valencia et al.~2007 or Figure 3 of Winn et al.~2011).
Super-Earths have surface escape velocities that, at $v_{\rm esc,p} \simeq 25$ km/s, are almost certainly too large for gas that is heated to temperatures 
$T \simeq 2000$--3000 K to escape at substantial rates
via Parker-type winds. The efficiency of driving a thermal
wind drops rapidly with increasing $v_{\rm esc,p}$.
The {\it local} hydrostatic scale height of gas (valid for altitudes
$h \lesssim R_p$)
\begin{equation} \label{eqn:h}
h_{\rm scl} \sim \frac{2c_s^2}{v_{\rm esc}^2} R_{\rm p} \sim 2 \times 10^{-3} \left( \frac{30}{\mu} \right) \left( \frac{10 M_{\oplus}}{M_p} \right) \left( \frac{R_p}{2 R_{\oplus}} \right) R_{\rm p} 
\end{equation}
decreases as $1/v_{\rm esc}^2$. For super-Earths, the base of a
Parker-type wind (the base is roughly identified as that location
where the local value of $v_{\rm esc}/c_s \sim 6$) would lie several
$R_p$ away from the planetary surface, while for a super-Mercury the
base would lie close to the surface.  Because the gas density must
decrease more or less according to $\rho_{\rm g} \propto
\exp{(-h/h_{\rm scl})}$ from the surface to the base, the gas density at
the base on a super-Earth would be many orders of magnitude lower than
on a super-Mercury.

Alternate mechanisms of mass loss from super-Earths, other than
a Parker-type wind, include: (i) XUV photoevaporation (by analogy with hot
Jupiters), (ii) stellar wind drag, and (iii) stellar radiation
pressure.  All fail to explain the inferred mass loss rates for
KIC 1255b. The intercepted power from the stellar XUV luminosity is
at least a factor of 10 too low compared to $L_{\rm required}$ (see
equation \ref{eqn:Lreq}), even for optimistically high values of
$L_{\rm XUV,s} = 10^{-5} L_{\odot}$ and of the efficiency $\varepsilon
= 0.1$ with which XUV photon energy is converted to mechanical
work.\footnote{XUV luminosities of K dwarfs are typically 3 $\times$
  greater than those of G dwarfs (Hodgkin \& Pyle 1994; Lecavelier des
  Etangs 2007).}$^,$\footnote{The efficiency $\varepsilon$ is lowered
  by ionization and radiative losses; radiative losses may be
  especially severe in the case of KIC 1255b because of cooling by
  dust, a problem that does not afflict hot Jupiter winds.}
Stellar wind drag suffers the same problem that XUV photoevaporation
does: the power intercepted by the planet from the stellar wind is
insufficient to supply $L_{\rm required}$.  An easy way to see this is
to note that the mechanical luminosity of the Solar wind is $L_{\rm
  wind,s} \sim (1/2) \dot{M}_{\odot} v_{\rm wind,s}^2 \sim (1/2) \cdot
(2 \times 10^{-14} M_{\odot}/{\rm yr}) \cdot (400 \,{\rm km/s})^2 \sim
3 \times 10^{-7} L_{\odot} < L_{\rm XUV,s}$.\footnote{Mass loss rates
  of K stars are observed to be comparable to that of the Sun (Wood et
  al.~2005; Cranmer 2008).}$^,$\footnote{At the orbital distance of
  KIC 1255b, the energy carried by the stellar wind is mostly magnetic
  and thermal, not kinetic. But after accounting for all three forms
  of energy using empirical measurements of the Solar wind at $a = 2.8
  R_{\odot}$, we arrived at roughly the same total power output,
  $L_{\rm wind,s} \sim 10^{-6} L_{\odot}$.} The intercepted power is
still too low even were we to account for a planetary magnetosphere,
which would increase the cross section presented by the planet by a
factor $\lesssim 10$.\footnote{If KIC 1255b has a surface dipole field
  of 1 G, then its magnetospheric radius is $\sim$$2 R_{\rm p}$.\label{foot:magneto}}
Finally, based on the
calculations of Kimura et al.~(2002), the force of stellar radiation
pressure acting on a grain is at most $\sim$30\% of the force of
planetary gravity on a super-Earth, for conditions in KIC 12557548
(but see \S\S\ref{sec:dustflow}--\ref{sec:scatt}).  
Moreover, even if the force of radiation pressure on a
grain were somewhat larger than that of planetary gravity, the grain
would be dragged by ambient gas, and might never accelerate to escape
velocity.

For all these reasons, super-Earths cannot launch dusty outflows like those
we infer for KIC 12557548---whereas sub-Earths can. There is still another
reason why dust clouds are especially observable for KIC 12557548
and not in the other observed systems.
Even if dust grains could be driven off the hot super-Earths detected to date,
they would be too hot; grains in these other systems would
sublimate too quickly to travel far from their parent planets,
and would fail to block as much light from their host stars.
If we use equation (\ref{eqn:teff}) to estimate the
temperature of a grain at the location of each super-Earth, we find:
\begin{itemize}
\item Kepler-10b : 3000 K
\item 55 Cnc e : 2800 K
\item CoRoT-7b : 2500 K 
\item KIC 1255b : 2100 K
\end{itemize}
Thus KIC 1255b---orbiting the coolest star---is distinguished in
having the coolest grains.  Sublimation lifetimes $t_{\rm sub}$ are
exponentially sensitive to temperature.  For example, whereas we have
estimated $t_{\rm sub} \sim 3 \times 10^4$ s for 0.2-$\mu$m-sized pyroxene
grains emitted by KIC 1255b, these same grains emitted by CoRoT-7b
would have $t_{\rm sub} \sim 10^2$ s (Kimura et
al.~2002)---considerably shorter than the travel time $t_{\rm travel}
\sim 2 \times 10^4$ s required to produce a $\sim$1\% occultation
(\S\ref{sec:dustcomp}).

Thus KIC 1255b apparently occupies a sweet spot for the production of occulting
dust clouds: the system is hot enough for the surface to vaporize and for a wind
to be launched,\footnote{
The three sub-Earths orbiting KOI-961 (Muirhead et al.~2012) are probably too cold to satisfy this condition.}
but cool enough that dust particles can be
formed from certain minerals not quite hot enough to vaporize (or to
form by condensation upon cooling of the vapor after it flows to a
different location), and to allow these dust particles to travel a
significant fraction of the host stellar radius.
 
\section{Summary and Outlook}
\label{sec:summary}

We have reported on occultations of KIC 12557548 that recur with a
period of 15.6854 hours, have depths that range from 1.3\% to
$\lesssim 0.15$\%, and that vary on timescales comparable to, if not
also shorter than, the occultation period. A search for periodic
modulation in the depths of the occultations did not reveal anything
definitive.  From various lines of evidence,
most notably a newly acquired optical spectrum, the target star appears
to be a mid-K dwarf.

We briefly considered a scenario wherein the changing transit depths
are due to a dual giant planet system (either on separate orbits or in
a binary pair) in which one of the planets undergoes grazing transits of the
target K star.  The orbital plane of the transiting planet would have to
precess in such a way as to effect the variable occultation depths.
We find, however, that a binary planet configuration is very likely to
be dynamically unstable, even if the two planets are themselves
essentially in a contact configuration.  In the case of two planets on
separate orbits, it is not clear how orbital precession could explain
either the remarkable changes from orbit to orbit or the
long intervals of $\sim$10 and 20 days when the occultations mostly
disappear.  We also briefly considered an eclipsing binary 
that is orbiting KIC 12557548 in a hierarchical triple configuration, but were unable to explain the basic properties of the observed
light curve in the context of such a scenario. 

We come down in favor of a scenario in which dust is aerodynamically
dragged off an orbiting rocky planet---possibly one not much larger
than Mercury in size---by a thermal Parker-type wind composed of metal
atoms sublimated off the planet's surface at a temperature of
$\sim$2000 K. 
The atmosphere may be loaded with dust either through condensation
into clouds, or through explosive volcanic activity.
The mass loss rate must sometimes be as high as
a few times $10^{11}$ g/s, or about $1 \,M_{\oplus}/$Gyr, and may comprise
by mass roughly equal parts gas and dust. Our fiducial
mass and radius for the underlying planet are $0.1 \,M_{\oplus}$ and
$0.5 \,R_{\oplus}$, respectively. The corresponding lifetime against
evaporation is $\sim$0.2 Gyr. This is probably less than the age of
the star, but is not alarmingly short.  Because the mass loss rates are
inferred from the observations and as such are largely fixed, planets
much smaller in mass than our fiducial super-Mercury have implausibly
short evaporation lifetimes, while planets much larger in mass and
thus in surface gravity do not seem capable of sustaining the required
mass loss rates.

After leaving the planet's gravitational well, the dusty gas will soon
become rarified to the point that the dust and gas trajectories
diverge.  We have simulated how dust grains, decoupled from gas, would
be shaped into a comet-like tail by the Coriolis force and stellar
radiation pressure. The head-tail asymmetry of the dust cloud promises
to explain the observed ingress-egress asymmetry in the occultation
profile. Pyroxene grains, each a fraction of a micron in size, are a
good candidate for the obscuring grains because of their relatively
long lifetimes against sublimation---alumina is also a promising
high-temperature mineral to consider.  Time variability of the outflow's
dust density as reflected in time variability of the occultation
depths could be caused by any number of factors: irregularities and
uneven outgassing rates on the planet's possibly volcanic surface; a
feedback loop between mass loss and the stellar radiation flux
received at the planet's surface that causes both to fluctuate with
time; and instabilities resulting from the interaction of the ionized
planetary wind with the magnetized stellar wind. The analogy between
our evaporating super-Mercury and comets may not be a bad one,
but much theoretical work remains to be done to put this scenario
on a more quantitative and secure footing.

\subsection{Observational Prospects}

Obviously, short-cadence observations with {\em Kepler} would be
helpful in clarifying the issue of whether there are shape changes in
individual occultations vs.~only changes in the overall depths. 
Some shape changes would be expected, as the timescale for the wind to
``refresh'' is given by the grain travel time $t_{\rm travel} \sim 5$
hr (equation \ref{eqn:t_travel}), which is not much longer than the
eclipse duration of 1.5 hr.
The target star, at $K_p$ = 15.7
magnitudes, is not very bright, but each 1-minute sample with {\em
  Kepler} would yield $1\,\sigma \sim 0.15$\% counting statistics.
Given that the depths of the deepest occultations are 1.3\%, this
would provide a signal-to-noise of 8:1 for individual samples. 
The ingress and egress should be resolved in short-cadence data and the
shapes should give important clues as to the nature of the system.

Observations with a large optical/infrared telescope, either from the
ground or from space, would also prove helpful. Measurement
of the wavelength dependence of the occultation depth could confirm
our hypothesis that sub-micron-sized dust is responsible.
The detection of solid-state absorption features
(e.g., the 10 $\mu$m silicate band) would also clinch the case for
dust.  
Deep imaging would have the added benefit of further ruling out
blends with faint, variable interlopers.

Searching for extra in-transit absorption by
metals---e.g., photoionized Mg, O, Si, Ca, and Fe---through
observations of spectral lines could
lead to a direct demonstration that the underlying
planet is comprised largely of heavy elements, and 
is evaporating (cf.~Linsky et al.~2010).  Our
rough calculations indicate that in some lines (e.g., the Mg II 2800
\AA ~doublet), the gaseous outflow may be so optically thick that its
occulting size is limited only by the confining pressure of the
incident stellar wind.  We estimate the corresponding maximum flux
decrement in any spectral lines to be on the order of $\sim$5\%.

If there indeed turns out to be a super-Mercury orbiting KIC
12557548, where the occultations are due to dust carried by outflowing
sublimated gas, the planet would be among the smallest-mass bodies
ever to have been (indirectly) detected. It would be the first extrasolar
planet shown to be geologically active and to be
disintegrating via the loss of high-Z material.

\acknowledgments
We thank the anonymous referee for an encouraging and incisive
report that motivated us to consider volcanic activity
and to examine more quantitatively the ability of the atmosphere
to entrain solids; Raymond Jeanloz and Michael Manga for
instructive exchanges about vaporizing silicates and volcano ejecta
speeds; Josh Carter for discussions about data validations and the
viability of a dynamically stable binary planet; 
Ruth Murray-Clay for input about thermal winds; Robert Szabo for information 
about properties of RR
Lyrae stars; and Bryce Croll, Dan Fabrycky, Ron Gilliland, Meredith
Hughes, John Johnson, Heather Knutson, Tim Morton, Margaret Pan, Erik
Petigura, and Josh Winn for stimulating discussions about follow-up
observations.  
We consulted with Ron Remillard, Rob Simcoe, and Adam
Burgasser about spectral classifications.  We would also like to thank
Robert Lamontagne and the staff at the Observatoire Astronomique du
Mont-M\'egantic for their assistance. EC is grateful for support from
the National Science Foundation, 
and for useful and encouraging
feedback from participants of the Berkeley Planet and Star Formation
Seminar, including Ryan O'Leary and Geoff Marcy who shared their own
analyses of the {\it Kepler} data on KIC 12557548.
LN thanks the
Natural Sciences and Engineering Research Council (NSERC) of Canada
for financial support. BK is grateful to the MIT Kavli Institute for
Astrophysics and Space Research for the hospitality they extended
during her visit and the support provided by the Turkish Council of
Higher Education.

\end{document}